\documentclass{emulateapj}

\usepackage{graphicx}
\usepackage{txfonts}
\usepackage{amssymb}
\usepackage[usenames,dvipsnames]{color}           % For color text: \color command
\newcommand{\BE}{\begin{equation}}
\newcommand{\EE}{\end{equation}}
\newcommand{\BA}{\begin{eqnarray}}
\newcommand{\EA}{\end{eqnarray}}

\renewcommand{\vec}[1]{{\mathbf #1}}

\newcommand{\rmd}{ {\ \mathrm d} }

\newcommand{\pder}[2]{ \frac{\partial #1}{\partial #2} }
\newcommand{\grad}{ {\bf \nabla } }
\newcommand{\curl}{ {\bf \nabla} \times}

\newcommand{\surf}{ {\mathcal S} }

\newcommand{\dl}{~{\mathrm d} \mathbf{l}}

\newcommand{\bb}{\vec B}
\newcommand{\bbp}{\vec B_p}
\newcommand{\jj}{ \vec j}
\newcommand{\rr}{ \vec r}

\newcommand{\uu}{ \vec u}

\newcommand{\eq}[1]{Equation~(\ref{eq:#1})} 
\newcommand{\eqs}[2]{Equations~(\ref{eq:#1}) and (\ref{eq:#2})} 
\newcommand{\eqss}[2]{Equations~(\ref{eq:#1} -- \ref{eq:#2})} 
\newcommand{\sect}[1]{Section~\ref{sec:#1}} 
\newcommand{\sects}[2]{Sections~\ref{sec:#1} and \ref{sec:#2}} 
 
\newcommand{\app}[1]{Appendix~\ref{app:#1}}

\newcommand{\tab}[1]{Table~\ref{tab:#1}}

\newcommand{\fig}[1]{Figure~\ref{fig:#1}}
\newcommand{\figs}[2]{Figures~\ref{fig:#1} and \ref{fig:#2}}

\newcommand{\cf}{\textit{cf.} }
\newcommand{\eg}{\textit{e.g.}, }
\newcommand{\ie}{\textit{i.e.}, }

\shorttitle{Net electric currents in active regions}
\shortauthors{Dalmasse et al.}

\begin{document} 
 
   \title{The origin of net electric currents in solar active regions}   
   
   \author{
	K. Dalmasse$^{1,2}$,
	G. Aulanier$^{2}$,
	P. D\'emoulin$^{2}$,
	B. Kliem$^{3}$,
	T. T\"or\"ok$^{4}$,
	E. Pariat$^{2}$
	}
		
   \affil{$1$ CISL/HAO, National Center for Atmospheric Research, 
                 P.O. Box 3000, Boulder, CO 80307-3000, USA
                 }
	\email{dalmasse@ucar.edu}
   \affil{$2$ LESIA, Observatoire de Paris, PSL Research University, CNRS, 
   		Sorbonne Universit\'es, UPMC Univ. Paris 06,\\ Univ. Paris-Diderot, 
		Sorbonne Paris Cit\'e, 5 place Jules Janssen, 92195 Meudon, France
		}
   \affil{$3$ Institut f\"ur Physik und Astronomie, Universit\"at 
                 Potsdam, Karl-Liebknecht-Str. 24-25, D-14476 Potsdam, 
                 Germany
                 }
   \affil{$4$ Predictive Science, Inc., 9990 Mesa Rim Road, 
                 Suite 170, San Diego, CA 92121, USA
                 }

% \abstract{}{}{}{}{} 
% 5 {} token are mandatory
 
   \begin{abstract}
There is a recurring question in solar physics about whether or not electric currents are 
neutralized in active regions (ARs). 
This question was recently revisited using three-dimensional (3D) magnetohydrodynamic 
(MHD) numerical simulations of magnetic flux emergence into the solar atmosphere. 
Such simulations showed that flux emergence can generate a substantial net current 
in ARs. Another source of AR currents are photospheric horizontal flows. Our aim is 
to determine the conditions for the occurrence of net vs. neutralized currents with 
this second mechanism. 
Using 3D MHD simulations, we systematically impose line-tied, quasi-static, photospheric 
twisting and shearing motions to a bipolar potential magnetic field. 
We find that such flows: (1) produce both {\it direct} and {\it return} currents, (2) induce 
very weak compression currents --- not observed in 2.5D ---  in the ambient field present 
in the close vicinity of the current-carrying field, and (3) can generate force-free magnetic 
fields with a net current. We demonstrate that neutralized currents are in general produced 
only in the absence of magnetic shear at the photospheric polarity inversion line --- a special 
condition rarely observed. 
We conclude that, as magnetic flux emergence, photospheric flows can build up net currents 
in the solar atmosphere, in agreement with recent observations. These results thus provide 
support for eruption models based on pre-eruption magnetic fields possessing a net coronal 
current.
   \end{abstract}   


%
%________________________________________________________________

\section{Introduction} \label{sec:S-Introduction}

%
% Flares/CMEs and electric currents
%
Current-carrying magnetic fields are an essential ingredient for the generation of 
flares and coronal mass ejections (CMEs) in the solar atmosphere 
\citep[\eg][]{Rust94,Schrijver05,Shibata11,Aulanier14}. Indeed, such non-potential 
magnetic fields store the free magnetic energy that powers these phenomena. 
What remains controversial, though, is whether and how a net electric current, 
here meant to be non-zero if integrated over {\it one} photospheric magnetic polarity, 
is formed in the source regions of these phenomena.

%
% Eruption models and neutralization
%
These questions arise from different theoretical arguments according to which electric 
currents should, or should not, be neutralized in active regions 
\citep[ARs; \eg][]{Melrose91,Parker96}. The answer to these questions may have critical 
consequences for several theoretical flare and CME models, as well as for pre-eruptive 
magnetic fields, developed from magnetic configurations containing a net current 
\citep[\eg][]{Low77,VanTend78,Martens87,Titov99,Kliem06,Demoulin10}. Indeed, full 
neutralization implies that there is no net current in an AR. In these circumstances, 
\cite{Forbes10} pointed out that the eruption mechanism of these models may be 
inhibited. Their relevance may therefore be questioned if ARs currents are in fact 
neutralized.

%
%  Definition of direct and return currents
%
The question of current neutralization in ARs derives from the fact that the current flowing 
in isolated, confined magnetic flux tubes consists of two parts: the so-called {\it direct} and 
{\it return} currents \citep{Melrose91,Parker96}. 
In magnetohydrodynamics (MHD), the direct (or main) currents refer to the electric currents 
that are expected from the chirality of a twisted/sheared magnetic flux tube. For a flux rope, 
the direct currents are flowing in the central part of the twisted flux tube while the return 
currents are flowing around them \citep[\eg see Figure 3 of][]{Melrose91}. These return 
currents shield the ambient magnetic field from the direct currents.

%
% The origin of electric currents in ARs
%
ARs currents are believed to be built up by two main mechanisms: (1) the emergence of 
current-carrying magnetic flux-tubes from the solar convection zone (CZ) into the corona 
\citep[\eg][]{Leka96,MorenoInsertis97,Longcope00,Cheung14}, and (2) the stressing of 
the coronal magnetic field by sub-photospheric and photospheric horizontal flows 
\citep[\eg][]{McClymont89,Melrose91,Klimchuk92,Torok03,Aulanier10}.

%
% The origin of neutralization
%
It has been argued that both mechanisms should in principle produce neutralized currents. 
Mechanism (1) is believed to be associated with the rising through the CZ of {\it confined} 
magnetic flux tubes \citep[\eg][]{Parker55,Fan09Review}. For instance, let us consider 
the simplified case of a twisted flux tube carrying an electric current, $I$, in cylindrical 
geometry. The confinement of the flux tube to a finite cross section of radius, $R$, requires 
that the total current it carries must vanish for $r > R$ (see \app{A-Currents-2.5D}). 
Such a twisted flux tube possesses a non-zero internal current-density, $\jj$. Therefore, 
the flux tube must contain a second type of currents which neutralizes the core/direct 
currents: \ie having the same total strength but flowing in the opposite direction (and often 
assumed to be a surface current). Based on the simplified assumption of full emergence of 
confined magnetic flux tubes, one may then expect that mechanism (1) would transport neutralized 
currents into the solar corona, thus generating a current-neutralized AR. As for mechanism (2), 
{\it localized} (sub)-photospheric horizontal flows transfer twist and shear to the magnetic field 
in a finite coronal volume. This field will typically inflate, inducing currents through compression 
also in the ambient magnetic field. Since the (sub)-photospheric driving volume is finite, one may 
expect that the changes in the coronal field also remain restricted to a finite volume. If true, 
a complete shielding of the generated currents, \ie neutralized currents, would be implied.

%
% Observational studies
%
Observationally, the normal/vertical component of the electric current density, $j_z$, can 
be derived by applying Amp\`ere's law to photospheric vector-magnetograms (see 
\eq{Eq-Ampere-Law}). Despite the various uncertainties and difficulties, the measurements 
of photospheric transverse magnetic fields are becoming more and more reliable 
\citep[\eg][]{Leka96,Metcalf06,Wiegelmann06,Gosain14}. For this reason, increasing 
attention has been paid to deriving the properties of currents in solar ARs, and to testing 
their degree of neutralization \citep[\eg][]{Wilkinson92,Leka99,Venkatakrishnan09,Sun12}. 
Some recent observational studies report the presence of direct and return currents in each 
magnetic polarity of ARs \citep[\eg][]{Wheatland00,Ravindra11,Georgoulis12,Gosain14}. 
These studies find both types of ARs, \ie some with neutralized currents, and some 
with a net current.

%
% Zeuh Problemeuh
%
The indications for the existence of ARs with a net current are at variance with theoretical 
arguments invoked in favor of current neutralization. Considering their past and present 
limitations, the relevance of the observational measurements has thus been questioned 
(\eg \citealt{Parker96}; see also the Introduction of \citealt{Georgoulis12}). On the other hand, 
the arguments in favor of current neutralization may well be oversimplified. For instance, 
they usually do not consider possible effects that become relevant in a fully three-dimensional 
(3D) geometry or in the case of partial magnetic flux emergence. This has thus led to a long-lasting 
debate about whether or not a net current can exist in ARs \citep[see \eg][]{Melrose91,Melrose95,Parker96}.

%
% Neutralization in MHD simulations
%
Numerical MHD simulations provide a useful alternative for addressing this problem. 
However, the neutralization of electric currents has barely been analyzed with this tool. 
A few MHD simulations reported the presence of both direct and return currents generated 
by photospheric line-tied motions applied to initially potential coronal fields 
\citep[\eg][]{Aulanier05,Delannee08}. Yet, only \cite{Torok03} quantified the associated degree 
of current neutralization. For the case of a twisted flux tube, they found that the neutralization 
only occurs when the photospheric motions do not extend to the polarity inversion line (PIL), 
so that no magnetic shear is built up at the initially unsheared PIL.

%
% Torok14
%
\cite{Torok14} were the first to revisit the question of current neutralization by means 
of 3D MHD simulations of magnetic flux emergence. In their experiment, the authors 
modeled the emergence of a buoyant magnetic flux rope into a stratified, plane-parallel 
atmosphere in hydrostatic equilibrium \citep[see][]{Leake13}. For that purpose, 
a sub-photospheric magnetic flux rope containing neutralized currents was considered. 
It was found that a complex redistribution of the initially sub-photospheric direct and 
return currents occurs in the vicinity of the photosphere. This subtle redistribution led 
mainly to the emergence of the initial direct currents \citep[see Figure 3b of][]{Torok14}, 
causing the development of a strong net current in the corona. This net current was 
associated with the development of a strong magnetic shear along the PIL, and 
some non-force-free return currents \citep[see Figure 5 of][]{Torok14}. These results 
indicate that the emergence of a current-neutralized magnetic flux tube can lead 
to the generation of a net current in ARs, as suggested by \cite{Longcope00}.

%
% Goal and outlines of the paper
%
In the present study, we pursue the work initiated by \cite{Torok14}, analyzing here 
the distribution and neutralization of currents generated by photospheric horizontal 
flows. We perform a parametric analysis of 3D, zero-$\beta$, MHD simulations 
of photospheric twisting and shearing motions imposed on a bipolar potential field. 
\sect{S-MHD} describes the main set-up of 
our numerical models. The results of our parametric study are presented 
in \sect{S-Twisting-Results} for the twisting motions and in \sect{S-Shearing-Results} 
for the shearing ones. A discussion and interpretation of our results is provided 
in \sect{S-Discussion}. Our conclusions are summarized in \sect{S-Conclusions}.

%%%%%%%%%%%%%%%%%%%%%%%%%%%%%%%%%%%%%%%%
%%%%%%%%%%%%%%%%%%%%%%%%%%%%%%%%%%%%%%%%
%%%%%%%%%%%%%%%%%%%%%%%%%%%%%%%%%%%%%%%%
%%%%%%%%%%%%%%%%%%%%%%%%%%%%%%%%%%%%%%%%

\section{Numerical model} \label{sec:S-MHD}

\subsection{Equations, numerical scheme, and boundary conditions} \label{sec:S-MHD-eq}

The numerical simulations described in this paper were performed in cartesian 
geometry, using the {\it Observationally-driven High-order scheme Magnetohydrodynamic} 
code \citep[OHM; see][]{Aulanier05}. We use the code in its zero-$\beta$ version 
in which the mass density, $\rho$, the fluid velocity, $\uu$, and the magnetic field, 
$\bb$, are advanced in time according to
	\BA   \label{eq:Eq-MHD-equations}
		\pder{\rho}{t} & = & - \grad \cdot (\rho \uu) \,, \\		
			\label{eq:Eq-FPD}
		\pder{\uu}{t} & = & - \left( \uu \cdot \grad \right) \uu + \left( \curl \bb \right) \times \bb / (\mu_0 \rho) + \tilde{\nu} \tilde{\Delta} \uu  \,,  \\
			\label{eq:Eq-Induction}			
		\pder{\bb}{t} & = & \curl \left( \uu \times \bb \right) + \eta \Delta \bb   \,,
	\EA
where $\tilde{\nu}$ and $\eta$ are diffusion coefficients. $\tilde{\nu}$ is a pseudo-viscosity, 
and $\eta$ is the electrical resistivity. These diffusions coefficients are used to limit 
the development of sharp discontinuities that may develop at the scale of the mesh 
and lead to quickly-growing numerical instabilities \citep[see][]{Aulanier05,Janvier13}. 
The solenoidal condition ($\grad \cdot \bb =0$; a discussion on its very weak magnitude 
is provided in \app{A-Solenoidal-Condition}) and the current density are not calculated 
in the code. The latter is derived from Amp\`ere's law
	\BE   \label{eq:Eq-Ampere-Law}
		\jj = \frac{1}{\mu_0} \curl \bb   \,.
	\EE

\eqss{Eq-MHD-equations}{Eq-Induction} are solved in non-dimensionalized units, using 
$\mu_0 = 1$. The velocities are expressed in units of the initially uniform Alfv\'en speed, 
$c_A (t=0) = 1$. The time unit is given by the Alfv\'en time, $t_A = 1$, which corresponds 
to the travel time of an Alfv\'en wave over a distance $d=1$. The diffusion coefficients are 
prescribed in terms of uniform characteristic speeds, $u_{\tilde{\nu}}$ and $u_{\eta}$, 
such that $\tilde{\nu} = u_{\tilde{\nu}} / l_{i,j,k}$ and $\eta = u_{\eta} l_{i,j,k}$, where $l_{i,j,k}$ is 
the smallest grid-spacing at point $(x_i ; y_j ; z_k)$. 
For all simulations, we set $u_{\tilde{\nu}} = 15 \% c_A (t=0)$ and $u_{\eta} = 1.5 \% c_A (t=0)$ 
in the coronal volume (the diffusion parameters are set to zero at the photospheric line-tied 
boundary).

All simulations are performed in the same domain covering 
$x \times y \times z \approx [-9.1,9.1]^2 \times [0,30]$, discretized on an non-structured 
mesh of $n_x \times n_y \times n_z = 231^3$ points. In the sub-domain $x \times y = [-1.5,1.5]^2$, 
the mesh is set uniform along the $x$ and $y$ directions, with mesh intervals 
$d^{\mathrm{min}}_x=d^{\mathrm{min}}_y=0.02$. 
This choice allows us to employ the higher mesh-resolution in the region where the strongest 
gradients of electric currents are generated. Outside of this region, the mesh is non-uniform 
in both the $x$ and $y$ directions, with mesh intervals defined such that 
$d^{i+1}_{x} / d^{i}_{x}=d^{j+1}_{y} / d^{j}_{y}=1.091$. The mesh is non-uniform 
all along the $z$ direction, with mesh intervals defined by $d^{k+1}_{z} / d^{k}_{z}=1.013$ 
and $d^{\mathrm{min}}_{z}=0.02$. With these settings, the largest mesh intervals are 
$d^{\mathrm{max}}_{z}=0.41$ along $z$ and $d^{\mathrm{max}}_{x,y}=0.65$ along 
both the $x$ and $y$ directions.

\begin{figure}
   \centering
  	 \centerline{\includegraphics[width=0.49\textwidth,clip=]{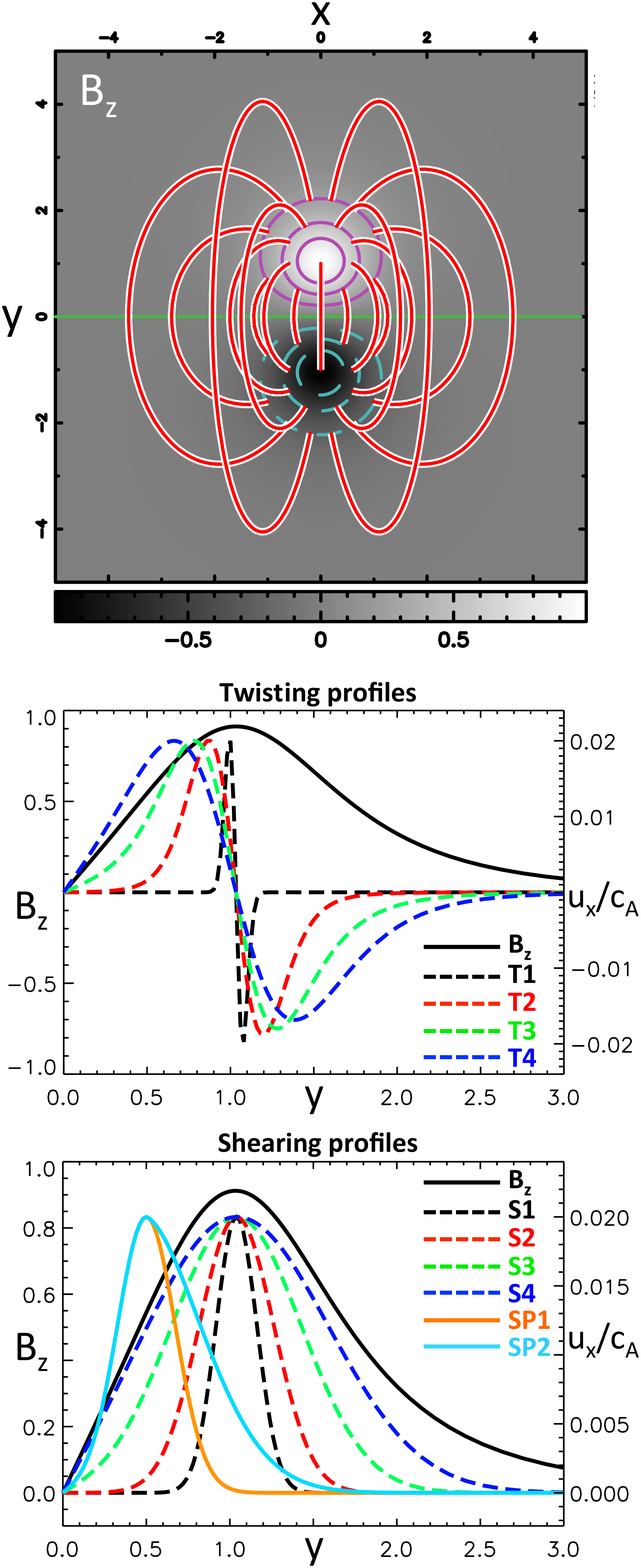}
	 }
   \caption{Top: initial magnetic field configuration of the shearing and twisting simulations. The synthetic magnetogram is represented at the $z=0$ plane with superposed $\pm [0.25;0.50;0.75]$ isocontours of $B_z$ (solid purple and dashed cyan) and selected field lines (red). The green line shows the polarity inversion line. Center: photospheric ($z=0$) $B_z$ and $u_x$ profiles in the $y$ direction for the twisting motions (see \sect{S-Twisting-Profile}). Bottom: same but for the shearing motions (see \sect{S-Shearing-Profile}).}
              \label{fig:Fig-BzVel-profiles}
    \end{figure}

We use the same initial conditions for each numerical simulation with an initially potential 
magnetic field constructed by placing two fictitious opposite magnetic charges below 
the photosphere. The positive and negative charges are placed at $\rr_{\pm} = (0,\pm1,-1)$ 
respectively. The resulting potential field is
	\BE   \label{eq:Eq-Initial-Bfield}
		\bbp (\rr) = q_0 \frac{\rr - \rr_{+}}{|\rr - \rr_{+}|^{3}} - q_0 \frac{\rr - \rr_{-}}{|\rr - \rr_{-}|^{3}}   \,,
	\EE
where $q_0 (> 0)$ is the strength of the two magnetic charges. For all simulations, 
we use $q_0 = 1$. The corresponding magnetic field is displayed in \fig{Fig-BzVel-profiles}. 
The initial plasma density is set to $\rho_0 = \bbp^{2} / (\mu_0 c^{2}_{A})$, where $c_{A}$ 
is the initially uniform Alfv\'en speed.

The top and all side boundaries of the numerical domain are open 
\citep[further details on these boundary conditions are given in Section 2.5 of][]{Aulanier05}. 
Line-tied boundary conditions are prescribed at the photospheric, or $z=0$, plane to build-up 
twisted and sheared magnetic fields.

%:          Figure: Bz with FLs + Bz & Velocity profiles
%%%%%%%%%%%%%%%%%%%%%%%%%%%%%%%%%%%%%%%%%
%%%%%%%%%% FIGURE ?? %%%%%%%%%%%%%%%%%%%%%%%%%%
%%%%%%%%%%%%%%%%%%%%%%%%%%%%%%%%%%%%%%%%%
%%%%%%%%%%%%%%%%%%%%%%%%%%%%%%%%%%%%%%%%%
%%%%%%%%%%%%%%%%%%%%%%%%%%%%%%%%%%%%%%%%%

%%%%%%%%%%%%%%%%%%%%%%%%%%%%%%%%%%%%%%%%
%%%%%%%%%%%%%%%%%%%%%%%%%%%%%%%%%%%%%%%%

\subsection{Photospheric twisting motions} \label{sec:S-Twisting-Profile}

%:          Figure: B3D for one of each Vprofile
%%%%%%%%%%%%%%%%%%%%%%%%%%%%%%%%%%%%%%%%%
%%%%%%%%%% FIGURE ?? %%%%%%%%%%%%%%%%%%%%%%%%%%
   \begin{figure*}
   \centering
  	 \centerline{\includegraphics[width=0.98\textwidth,clip=]{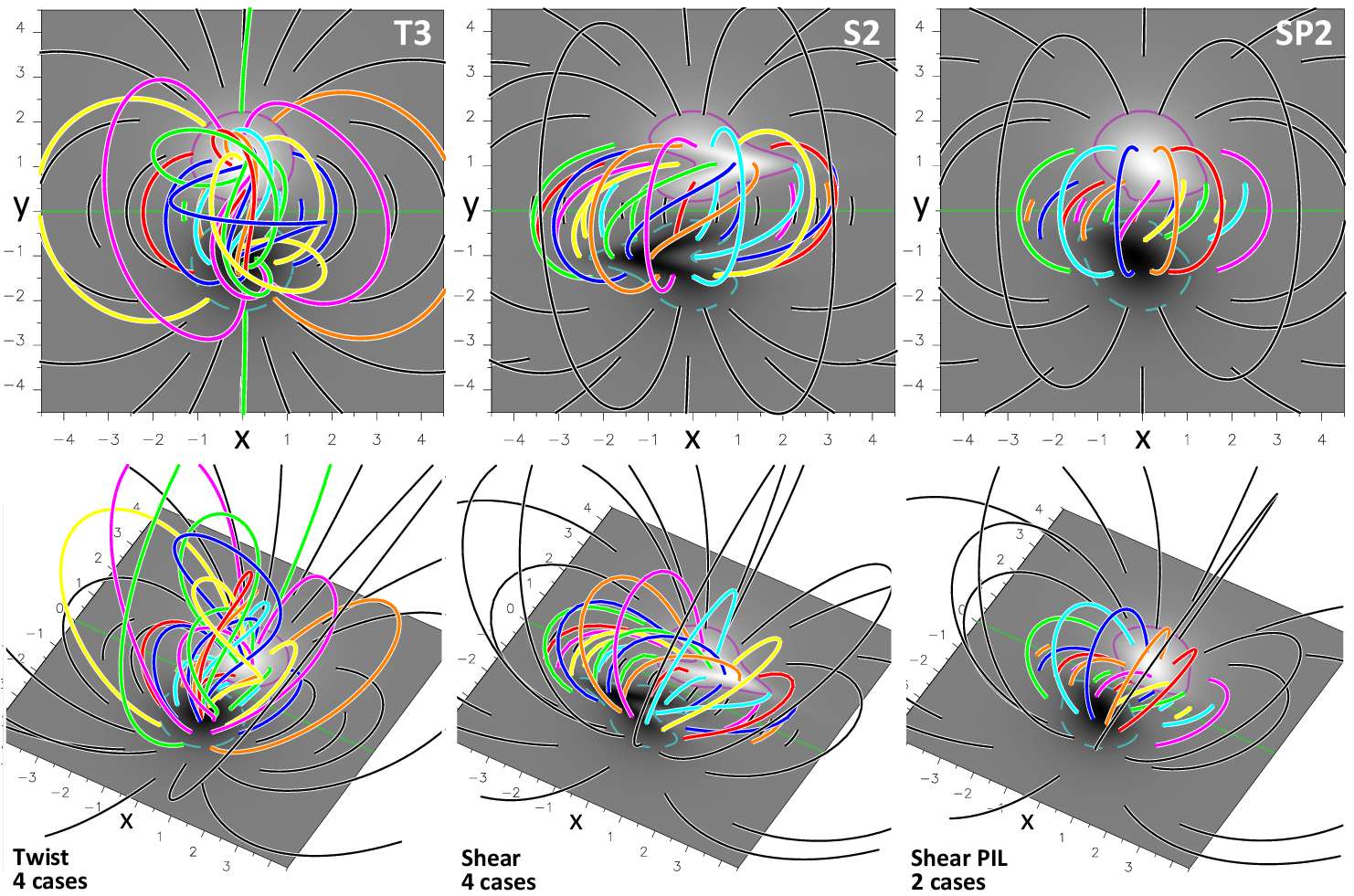}
	 }
   \caption{Top (top row) and 3D (bottom row) views of the magnetic field generated by the photospheric line-tied motions. Left: twist case $\mathrm{T}3$ (\eqss{Eq-Twisting-Vxprofile}{Eq-Twisting-Pprofile}). Center: shear case $\mathrm{S}2$ (\eq{Eq-Shearing-Vprofile1}). Right: shear case $\mathrm{SP}2$ (\eqs{Eq-Shearing-Vprofile2-LEY1}{Eq-Shearing-Vprofile2-GEY1}). Current-carrying (potential) magnetic field lines are in color (black).}
              \label{fig:Fig-B3D}
    \end{figure*}
%%%%%%%%%%%%%%%%%%%%%%%%%%%%%%%%%%%%%%%%%
%%%%%%%%%%%%%%%%%%%%%%%%%%%%%%%%%%%%%%%%%
%%%%%%%%%%%%%%%%%%%%%%%%%%%%%%%%%%%%%%%%%

The first type of photospheric driving is applied to twist the initial potential 
magnetic configuration along the isocontours of the photospheric vertical magnetic 
field, $B_z (z=0)$. This modifies the transverse components of the initial potential field 
while preserving its photospheric flux distribution, $B_z (z=0,t=0)$. The expression of 
the twisting velocity field is
	\BA   \label{eq:Eq-Twisting-Vxprofile}
		u_{x} (z=0) & = & u_0 \pder{\psi}{y}  \,,  \\
		u_{y} (z=0) & = & - u_0 \pder{\psi}{x}  \,,  \\
		\label{eq:Eq-Twisting-Vyprofile}
		u_{z} (z=0) & = & 0   \,,
	\EA
where $u_0$ is a free parameter that controls the maximum speed of the driving, and 
$\psi$ is a normalized, time-dependent potential \citep[see \eg][]{Amari96,Torok03,Aulanier05}. 
This potential depends on $B_z (z=0)$ such that
	\BE   \label{eq:Eq-Twisting-Pprofile}
		\psi = \frac{B^{2}_{z} (z=0)}{B^{\mathrm{max} \ 2}_{z} (z=0)} \exp \left( \frac{B^{2}_{z} (z=0) - B^{\mathrm{max} \ 2}_{z} (z=0)}{ \zeta^{2}_{tw} B^{\mathrm{max} \ 2}_{z} (z=0)} \right)   \,,
	\EE
This twisting profile generates two vortices centered on $\pm B^{\mathrm{max}}_{z} (z=0)$ 
respectively. The size of the vortices is controlled by the free-parameter $\zeta_{tw}$.

All twisting profiles were applied with a maximum driving speed $u_0 = 0.02 c_A (t=0)$. 
\fig{Fig-BzVel-profiles} shows the profiles of $u_x (x=0,y)$ for all four cases considered 
in our study, and referred to as $\mathrm{T}\{1;2;3;4\}$. The corresponding parameters 
are listed in \tab{Parameters-Velocity-Profiles}. The left panels of \fig{Fig-B3D} display 
the 3D distribution of the magnetic field for $\mathrm{T}3$.

While the twisting boundary driving analytically preserves the initial distribution of 
the photospheric vertical magnetic field, small numerical errors eventually deform it 
on the long run. To ensure that $B_z$ is preserved in time for the twisting simulations, 
we numerically reset $B_z (z=0,t)$ to $B_z (z=0,t=0)$ at each time step in the course 
of the runs.

%:          Table: Parameters for the velocity profiles
%%%%%%%%%%%%%%%%%%%%%%%%%%%%%%%%%%%%%%%%%
%%%%%%%%%%%%%%%%%% Table ?? %%%%%%%%%%%%%%%%%
   \begin{deluxetable}{c c c c c c c c}
   \tablecaption{Values of the parameters for the velocity profiles
   	\label{tab:Parameters-Velocity-Profiles}
	}
   \tablehead{\colhead{Run} & \colhead{$\zeta_{tw}$} & \colhead{$\zeta_{sh}$} & \colhead{$\zeta_{sh1}$} & \colhead{$\zeta_{sh2}$} & \colhead{$y_0$} & \colhead{$y_1$} & \colhead{$y_2$}}
   \tablecomments{All values are in non-dimensional units (normalized by half the distance between the two photospheric magnetic polarities).}
   \startdata
	$\mathrm{T}1$ & 0.1 & - & - & - & 1 & - & -  \\
	$\mathrm{T}2$ & 0.5 & - & - & - & 1 & - & -  \\
	$\mathrm{T}3$ & 1 & - & - & - & 1 & - & -  \\
	$\mathrm{T}4$ & 5 & - & - & - & 1 & - & -  \\
	$\mathrm{S}1$ & - & 0.17 & - & - & 1 & - & -  \\
	$\mathrm{S}2$ & - & 0.32 & - & - & 1 & - & -  \\
	$\mathrm{S}3$ & - & 0.55 & - & - & 1 & - & -  \\
	$\mathrm{S}4$ & - & 0.77 & - & - & 1 & - & -  \\
	$\mathrm{SP}1$ & - & - & 0.24 & 0.24 & - & 0.5 & 0.5  \\
	$\mathrm{SP}2$ & - & - & 0.24 & 0.63 & - & 0.5 & 0.3
   \enddata
   \end{deluxetable} 
%%%%%%%%%%%%%%%%%%%%%%%%%%%%%%%%%%%%%%%%%
%%%%%%%%%%%%%%%%%%%%%%%%%%%%%%%%%%%%%%%%%

%%%%%%%%%%%%%%%%%%%%%%%%%%%%%%%%%%%%%%%%
%%%%%%%%%%%%%%%%%%%%%%%%%%%%%%%%%%%%%%%%

\subsection{Photospheric shearing motions} \label{sec:S-Shearing-Profile}

We consider two types of photospheric shearing motions along the $x$ direction, 
$\uu (z=0) = (u_x, 0, 0)$. The first is centered on the strongest magnetic field of both 
photospheric magnetic polarities. The second is confined to the weak field surrounding 
the PIL to primarily shear it \citep[\eg similarly to][]{Antiochos94}.

The first type of shearing motions (curves $\mathrm{S}\{1;2;3;4\}$ in \fig{Fig-BzVel-profiles}) 
is given by
	\BE   \label{eq:Eq-Shearing-Vprofile1}
		u_{x} (z=0) = u_0 \ a_0 \left[ \exp \left( 
		            - \frac{(y-y_0)^2}{\zeta^{2}_{sh}} \right)  
		            - \exp \left( - \frac{(y+y_0)^2}{\zeta^{2}_{sh}}   \right) \right]   \,.
	\EE
where $\zeta_{sh}$ controls the width of the shearing, $y_0$ corresponds to the $y$-position 
of the maximum velocity (shearing center), and $a_0$ is a constant used for normalization of 
the Gaussian-profile. The resulting shearing is invariant along the $x$ direction.

The second type of shearing motions (curves $\mathrm{SP}\{1;2\}$ in \fig{Fig-BzVel-profiles}) 
is given by
	\BE   \label{eq:Eq-Shearing-Vprofile2-LEY1}
		u_{x} (z=0) = u_0 \ a_1 \left[ \exp \left( 
		            - \frac{(y-y_1)^2}{\zeta^{2}_{sh1}} \right)  
		            - \exp \left( - \frac{(y+y_1)^2}{\zeta^{2}_{sh1}}    \right) \right]   \,,
	\EE
for $|y| \le y_1$, and 
	\BE   \label{eq:Eq-Shearing-Vprofile2-GEY1}
		u_{x} (z=0) = u_0 \ a_2 \left[ \exp \left( - \frac{(y-y_2)^2}{\zeta^{2}_{sh2}} \right)  - \exp \left( - \frac{(y+y_2)^2}{\zeta^{2}_{sh2}} \right) \right]   \,,
	\EE
for $|y| \ge y_1$. Choosing $\zeta_{sh1} \ne \zeta_{sh2}$ allows one to have a broader 
sheared region either for $|y| < y_1$ or $|y| > y_1$. The parameter $y_2$ is computed 
to ensure the continuity of the shearing profile at $|y| = y_1$, where $u_x (|y| = y_1,z=0) = u_0$.

In this paper, we consider four shearing profiles for the first model, referred to as 
$\mathrm{S}\{1;2;3;4\}$, and two shearing profiles for the second model, referred to as 
$\mathrm{SP}1$ and $\mathrm{SP}2$. The corresponding parameters are listed 
in \tab{Parameters-Velocity-Profiles}.

As for the twisting simulations, all shearing profiles were applied with a maximum driving 
speed $u_0 = 0.02 c_A (t=0)$ to ensure a quasi-force-free evolution of the magnetic field. 
The applied profiles are displayed in \fig{Fig-BzVel-profiles}. The middle and left panels of 
\fig{Fig-B3D} display the 3D distribution of the magnetic field for $\mathrm{S}2$ and 
$\mathrm{SP}2$, respectively.

%%%%%%%%%%%%%%%%%%%%%%%%%%%%%%%%%%%%%%%%
%%%%%%%%%%%%%%%%%%%%%%%%%%%%%%%%%%%%%%%%

\subsection{Ramp function} \label{sec:S-Ramp-Function}

We apply all photospheric line-tied motions using a ramp function to smoothly 
bring the system from rest to a constant-velocity photospheric driving, such that
	\BE   \label{eq:Eq-Time-Velocity}
		\uu (x,y,z=0,t) = \gamma (t) \ \uu (x,y,z=0) \,,
	\EE
where $\uu (x,y,z=0)$ is defined in \sects{S-Twisting-Profile}{S-Shearing-Profile}. 
The ramp function, $\gamma (t)$, is given by
	\BE   \label{eq:Eq-Ramp-profile}
		\gamma (t) =  \frac{1}{2} \tanh \left( \frac{2(t-t_m)}{t_{hw}} \right) + \frac{1}{2}  \,,
	\EE
where $t_m$ corresponds to the time at which the middle of the ramp function is 
reached, and $t_{hw}$ corresponds to the half-width of the ramp time. In our runs, 
we fix $t_m = 15 t_A$ and $t_{hw} = 5 t_A$. With these settings, the photospheric 
driving starts after $\approx 10$ Alfv\'en times, allowing the system to reach a good 
numerical equilibrium before the acceleration begins. The constant-velocity 
photospheric driving is reached after $\approx 20$ Alfv\'en times.

%%%%%%%%%%%%%%%%%%%%%%%%%%%%%%%%%%%%%%%%
%%%%%%%%%%%%%%%%%%%%%%%%%%%%%%%%%%%%%%%%
%%%%%%%%%%%%%%%%%%%%%%%%%%%%%%%%%%%%%%%%
%%%%%%%%%%%%%%%%%%%%%%%%%%%%%%%%%%%%%%%%

\section{Photospheric currents induced by twisting motions} \label{sec:S-Twisting-Results}

In cylindrical geometry, any twisting motion based on a twist function that falls to zero 
at a finite radial distance should generate a twisted flux tube formed of a core of direct 
currents, fully surrounded by a shell of return currents that exactly neutralize the direct 
currents (see \app{A-Currents-2.5D}).

The transposition of these results to 3D geometry has often been used to argue that 
compact photospheric twisting motions should lead to the generation of fully neutralized 
currents \citep[\eg][]{Melrose91,Parker96}. However, such a transposition disregards 
the fact that a 3D magnetic flux tube may not systematically have a cylindrical analogue. 
It is therefore not obvious that the properties of electric currents expected from a simplified, 
cylindrical geometry may still hold in a more complex 3D one.

To address this issue, we analyze the electric currents generated by compact, photospheric 
twisting motions (see \sect{S-Twisting-Profile}).

%%%%%%%%%%%%%%%%%%%%%%%%%%%%%%%%%%%%%%%%
%%%%%%%%%%%%%%%%%%%%%%%%%%%%%%%%%%%%%%%%

\subsection{Photospheric distribution of vertical currents} \label{sec:S-Twisting-JzDistribution}

\begin{figure}
   \centering
  	 \centerline{\includegraphics[width=0.47\textwidth,clip=]{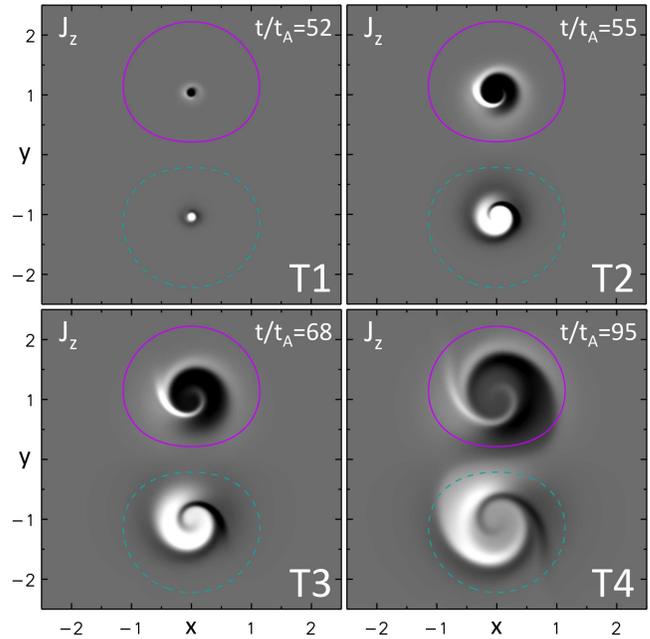}
	 }
   \caption{Photospheric ($z=0$) current maps, $j_z$, at the end of each simulations for the symmetric twisting of two opposite magnetic polarities (\sect{S-Twisting-Profile}), for $\mathrm{T}\{1;2;3;4\}$. White and black display positive and negative currents respectively. Values are saturated at $\pm 1.5$. Solid purple and dashed cyan lines represent $\pm 0.25$ isocontours of $B_z$.}
              \label{fig:Fig-Jztwist}
    \end{figure}

\fig{Fig-Jztwist} displays the photospheric vertical current density, $j_z$, 
for the twisting runs, a few Alfv\'en times before each numerical simulation terminated 
(due to the development of a numerical instability caused by very sharp gradients). 
Each twisting run induces the generation of a core of negative-direct currents 
surrounded by a shell of positive-return currents, just as in cylindrical geometry. 
As reported by \cite{Torok03} and \cite{Delannee08} for similar models, we find that 
the direct currents are overall stronger and more compact than the return currents. 
In addition, we notice that the distribution of currents always exhibits the same type 
of strong azimuthal asymmetry (except for $\mathrm{T}{1}$) that does not occur 
in cylindrical geometry.

This generic property was already explained by \cite{Aulanier05} and \cite{Titov08}, 
and is typical of the 3D loop-geometries analyzed in this paper. This asymmetry is 
an effect of field-line length resulting from the flux tube curvature. In cylindrical geometry 
$(r,\theta,z)$, the equations of magnetic field lines and electric current density lead to
	\BA   \label{eq:Eq-Btheta-Cylindrical}
		B_{\theta} & = & \frac{r \Phi}{l} B_z   \,,  \\
		\label{eq:Eq-J-Cylindrical}
		j_z & = & \frac{1}{r} \pder{r B_{\theta}}{r} \propto \frac{B_{\theta}}{r}  \propto \frac{\Phi B_z}{l}  \,,
	\EA
where $\Phi$ and $l$ are the field-line twist and length, respectively. For a given 
$B_z$ isocontour $\Phi$ is fixed. It then follows that shorter field lines (\ie smaller 
$l$) possess a stronger electric current density (\eq{Eq-J-Cylindrical}). For our 3D curved 
flux tube, the same amount of twist, $\Phi$, is transferred to each field line of any given 
isocontour of $B_z$. The previous considerations thus imply that stronger currents 
develop at the footpoints of the shortest field lines of any $B_z$ isocontour. This creates 
an asymmetry that is amplified by the faster expansion of the larger field lines. The advection 
of this asymmetry by the photospheric motions is responsible for the swirling pattern exhibited 
by the electric current distribution. This effect is extremely weak for $\mathrm{T}1$ because 
the twisting vortices are so narrow that the twisted field lines have a very similar length.

%:          Figure: Jz twisting
%%%%%%%%%%%%%%%%%%%%%%%%%%%%%%%%%%%%%%%%%
%%%%%%%%%% FIGURE ?? %%%%%%%%%%%%%%%%%%%%%%%%%%
%%%%%%%%%%%%%%%%%%%%%%%%%%%%%%%%%%%%%%%%%
%%%%%%%%%%%%%%%%%%%%%%%%%%%%%%%%%%%%%%%%%
%%%%%%%%%%%%%%%%%%%%%%%%%%%%%%%%%%%%%%%%%	

Among several differences between each simulation, one is the extension of the currents 
close to the PIL. In particular, for the cases $\mathrm{T}\{1;2\}$, the twisting vortices are 
so narrow that the distribution of current is mainly localized well inside the isocontour 
$| B_z | = 0.25$. On the contrary, for $\mathrm{T}\{3;4\}$, the vortices are so broad that 
the distribution of current extends to the PIL.

%%%%%%%%%%%%%%%%%%%%%%%%%%%%%%%%%%%%%%%%
%%%%%%%%%%%%%%%%%%%%%%%%%%%%%%%%%%%%%%%%

\subsection{Evolution of the total direct and return currents} \label{sec:S-Evolution-DRC-Twisting}

We now analyze the curves of integrated currents by computing
	\BE   \label{eq:Eq-Current}
		I^{\mathrm{X}}_{z} = \int_{y \ge 0} j^{\mathrm{X}}_{z} \rmd \surf   \,,
	\EE
where $X$ refers to the total {\it direct} or {\it return} current within the positive 
magnetic polarity (\ie $y \ge 0$). Because the direct/return currents are negative/positive, 
we compute the total direct/return current by extracting the negative/positive current 
density in the positive magnetic polarity.

\begin{figure}
   \centering	 
  	 \centerline{\includegraphics[width=0.47\textwidth,clip=]{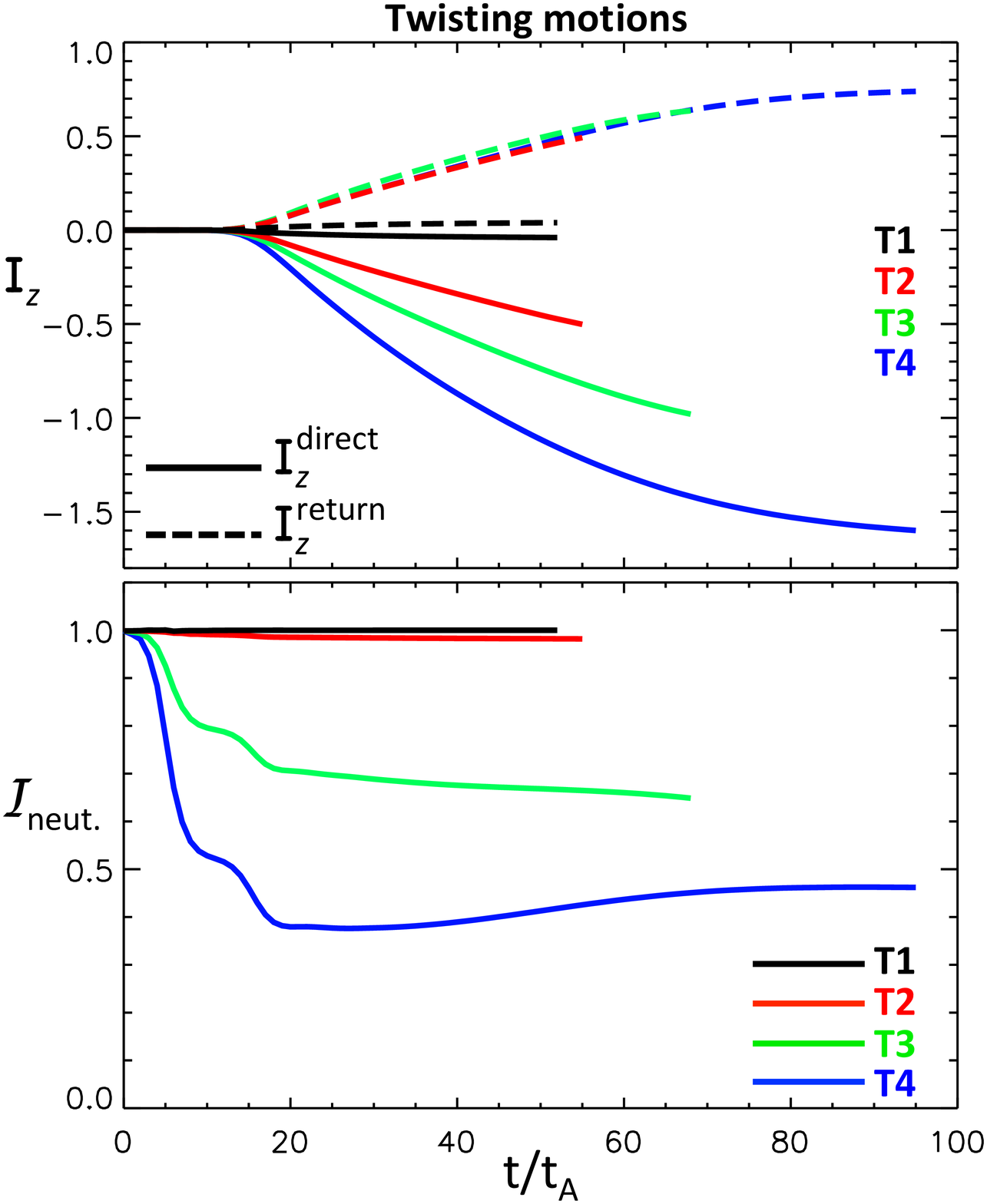}
	 }	 	 
   \caption{Evolution of the electric current, $I_z$, at $z=0$ in the positive polarity for the photospheric twisting motions. Top: negative-direct ($I^{\mathrm{direct}}_{z}$) and positive-return ($I^{\mathrm{return}}_{z}$) currents (respectively displayed in solid and dashed lines). Bottom: neutralization ratio.}
              \label{fig:Fig-Iztwist}
    \end{figure}

The temporal evolution of the integrated direct and return currents is presented 
in the top panel of \fig{Fig-Iztwist}. Due to its small vortex size, the $\mathrm{T}1$ case 
shows the development of only weak direct and return currents. For the three other cases, 
the vortices are broader and stronger direct and return currents develop.

%:          Figure: Iz & Neutralisation rate - twisting
%%%%%%%%%%%%%%%%%%%%%%%%%%%%%%%%%%%%%%%%%
%%%%%%%%%% FIGURE ?? %%%%%%%%%%%%%%%%%%%%%%%%%%
%%%%%%%%%%%%%%%%%%%%%%%%%%%%%%%%%%%%%%%%%
%%%%%%%%%%%%%%%%%%%%%%%%%%%%%%%%%%%%%%%%%
%%%%%%%%%%%%%%%%%%%%%%%%%%%%%%%%%%%%%%%%%

\fig{Fig-Iztwist} shows that the absolute value of the integrated direct current exhibit 
a monotonic rise (regardless of the vortex width) and the strength of the direct current 
increases with the width of the twisting vortex. The integrated return current also presents 
a monotonic behavior. However, the strength of the return current only increases with 
the vortex width up to $\zeta_{tw} \le 0.5$.

Increasing the vortex width builds up twist in a larger volume and generates 
a higher direct current. The return current is affected in a more complex manner, 
because the boundary between the direct and return currents is pushed closer 
to the PIL. Away from the PIL, increasing the vortex also implies that longer magnetic 
field lines are being twisted. As discussed by \cite{Aulanier05}, the resulting fast 
expansion of these field lines limits their current density.

We therefore conjecture that the increase of total electric current 
($| I^{\mathrm{direct}}_{z} | + | I^{\mathrm{return}}_{z} |$) with the vortex size, is caused 
by a complex competition of three mechanisms: (1) an increase due to the transfer 
of magnetic twist in a broader region, (2) a saturation/decrease caused by the approach 
of the line of current reversal to the PIL, and (3) a saturation/decrease induced by the fast 
expansion of the magnetic field lines.

%%%%%%%%%%%%%%%%%%%%%%%%%%%%%%%%%%%%%%%%
%%%%%%%%%%%%%%%%%%%%%%%%%%%%%%%%%%%%%%%%

\subsection{Evolution of the neutralization ratio} \label{sec:S-Evolution-NRate-Twisting}

In order to quantify the neutralization of electric currents generated by photospheric 
line-tied motions, we define the neutralization ratio, 
$\mathcal{I}_{\mathrm{neut.}}$, as
	\BE   \label{eq:Eq-Neutralization-Ratio}
		\mathcal{I}_{\mathrm{neut.}} = \bigg| \frac{I^{\mathrm{return}}_{z}}{I^{\mathrm{direct}}_{z}} \bigg|   \,,
	\EE
where $I^{\mathrm{direct}}_{z}$ and $I^{\mathrm{return}}_{z}$ are computed from 
\eq{Eq-Current}. The neutralization ratio is $1$ for fully neutralized currents and 0 
for a magnetic field solely containing direct currents.

The bottom panel of \fig{Fig-Iztwist} shows the temporal evolution of the neutralization 
ratio for the twisting runs. Our goal is to analyze the neutralization of currents for different 
spatial profiles of photospheric line-tied motions. In the following, we therefore restrict 
our analysis to the period during which the system evolves in response to a stationary 
boundary driving, \ie for $t \gtrsim 20 t_A$. A brief discussion of the results for the transition 
phase, $t \in [0;20] t_A$, is nonetheless provided in \app{A-Nratio-Transition-phase}.

We find that the electric currents remain fully neutralized ($\mathcal{I}_{\mathrm{neut.}} = 1$) 
during the run with the narrowest vortices ($\mathrm{T}1$), and nearly neutralized 
($\mathcal{I}_{\mathrm{neut.}} \approx 0.98$) for the $\mathrm{T}2$ run. On the contrary, 
the two runs with the larger vortices, $\mathrm{T}\{3;4\}$, exhibit a strong departure 
from neutralization. This confirms the earlier results of \cite{Torok03} that were obtained 
with different numerical settings.

\fig{Fig-Iztwist} further indicates that the constant boundary driving phase is associated 
with a slow and weak decrease of the neutralization ratio for run $\mathrm{T}3$. 
On the contrary, the neutralization ratio of run $\mathrm{T}4$ presents a weak 
increase followed by a saturation. Such behaviors result from a non-trivial combination 
of the increase of currents with magnetic twist and the saturation of currents caused 
by the fast expansion of the largest field lines, as discussed in \sect{S-Evolution-DRC-Twisting}. 
Note that for $\mathrm{T}4$, two additional effects are likely involved in the evolution 
of the neutralization ratio: (1) the twisted flux tube starts to leave the numerical domain 
(as suggested by the opening of some field lines, not shown here), and (2) the flux tube 
enters a super-exponential growth phase \citep[\cf the equilibrium curves in][]{Torok03,Aulanier05}, 
leading to a fast expansion of the more twisted field lines compared with the less twisted 
ones, \ie for the direct currents. This limits the growth of the direct current more efficiently 
than that of the return current, which could explain the observed weak increase of 
the neutralization ratio.

In contrast to cylindrical symmetry (\app{A-Currents-2.5D}), \fig{Fig-Iztwist} shows that twist 
profiles do not systematically generate neutralized currents in fully 3D twisted flux tubes. 
Hence, the condition for current neutralization derived in 2.5D geometry cannot be 
directly transposed to a fully 3D loop geometry such as the one considered in this paper. 
Indeed, we will show in \sect{S-Discussion} that (1) the condition for current neutralization 
in 3D is more subtle than in 2.5D, and (2) net currents are associated with 3D coronal fields 
that do not have any analogue in 2.5D.

%%%%%%%%%%%%%%%%%%%%%%%%%%%%%%%%%%%%%%%%
%%%%%%%%%%%%%%%%%%%%%%%%%%%%%%%%%%%%%%%%
%%%%%%%%%%%%%%%%%%%%%%%%%%%%%%%%%%%%%%%%
%%%%%%%%%%%%%%%%%%%%%%%%%%%%%%%%%%%%%%%%

\section{Photospheric currents induced by shearing motions} \label{sec:S-Shearing-Results}

In this section, we study the electric currents generated by compact photospheric 
shearing motions (see \sect{S-Shearing-Profile}).

%%%%%%%%%%%%%%%%%%%%%%%%%%%%%%%%%%%%%%%%
%%%%%%%%%%%%%%%%%%%%%%%%%%%%%%%%%%%%%%%%

\subsection{Photospheric distribution of vertical currents} \label{sec:S-Shearing-JzDistribution}

\begin{figure}
   \centering
  	 \centerline{\includegraphics[width=0.47\textwidth,clip=]{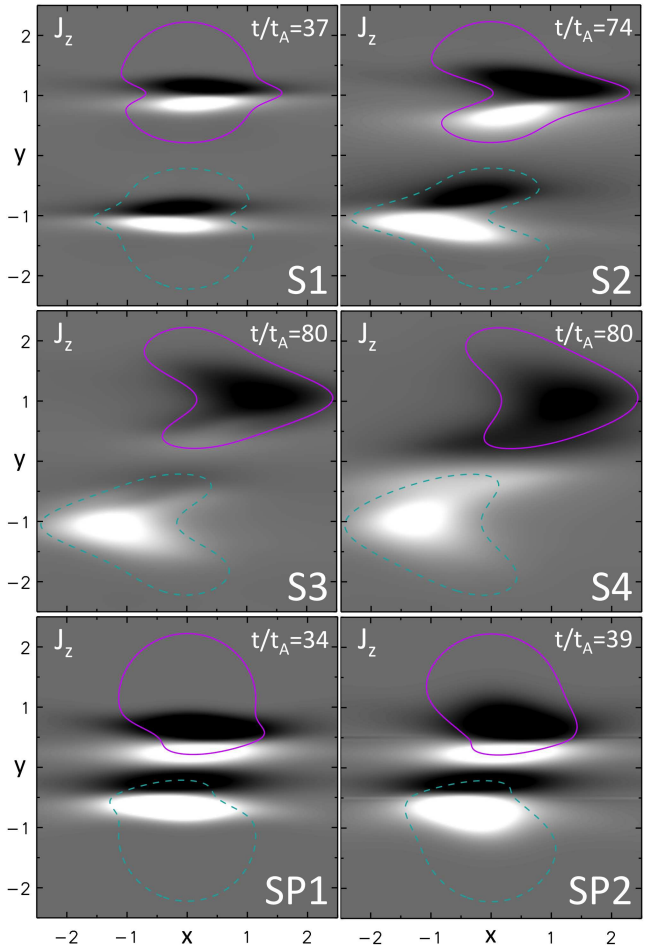}
	 }
   \caption{Photospheric current maps as in \fig{Fig-Jztwist} for the symmetric shearing of two opposite magnetic polarities (\sect{S-Shearing-Profile}). Top/middle: shearing centered on $B^{\mathrm{max}}_{z}$ for $\mathrm{S}\{1;2;3;4\}$. Bottom: shearing centered on the weak field surrounding the PIL for cases $\mathrm{SP}\{1;2\}$. The color coding is the same as in \fig{Fig-Jztwist} with saturation at $\pm 0.5$.}
              \label{fig:Fig-Jzshear}
    \end{figure}

\fig{Fig-Jzshear} displays the photospheric vertical current density, $j_z$, 
for the shearing runs. We find that each run presents both negative-direct and 
positive-return, force-free currents. Their sign are identified from the sign of 
magnetic helicity transferred to the system. In particular, the chosen shearing 
motions produce negative magnetic helicity, and hence, negative-direct currents.

For almost all simulations, the direct and return currents have a similar spatial 
distribution with similar intensities. Two cases, however, do not display the same 
pattern: $\mathrm{S}3$ and $\mathrm{S}4$ mostly possess direct currents. 
For these two runs, the return currents are much weaker than the direct ones 
(\eg $\approx 7$ times weaker for $\mathrm{S}3$).

%:          Figure: Jz shearing
%%%%%%%%%%%%%%%%%%%%%%%%%%%%%%%%%%%%%%%%%
%%%%%%%%%% FIGURE ?? %%%%%%%%%%%%%%%%%%%%%%%%%%
%%%%%%%%%%%%%%%%%%%%%%%%%%%%%%%%%%%%%%%%%
%%%%%%%%%%%%%%%%%%%%%%%%%%%%%%%%%%%%%%%%%
%%%%%%%%%%%%%%%%%%%%%%%%%%%%%%%%%%%%%%%%%

\begin{figure}
   \centering
  	 \centerline{\includegraphics[width=0.47\textwidth,clip=]{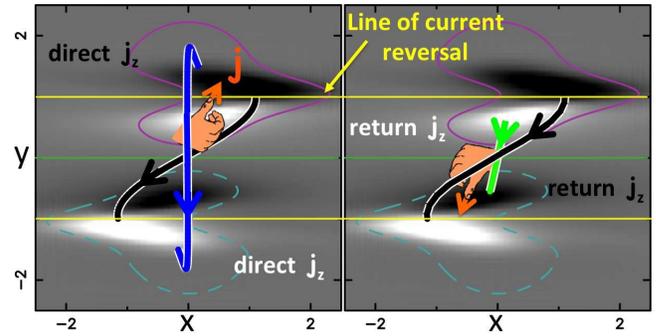}
	 }
   \caption{Examples of sheared magnetic field lines and current direction generated by photospheric shearing motions for the case $\mathrm{S}2$. The thick blue (respectively green) field line is anchored within the direct (respectively return) current. The thick black field line is anchored at the line of current reversal (thin yellow lines). The orange arrows and hands show the direction of current density, $\jj$, on both sides of the line of current reversal, as inferred from the {\it right-hand} rule. The arrows on the field lines indicate the magnetic field direction. The color coding and saturation are the same as in \fig{Fig-Jzshear}.}
              \label{fig:Fig-Jzshear-Sign}
    \end{figure}

The return currents are, in general, expected for localized shearing motions. This is 
qualitatively described in \fig{Fig-Jzshear-Sign} with the right-hand rule relating 
the local magnetic shear with the direction of the electric current. Localized shearing 
motions generate a curl of the magnetic field that changes sign at the line of strongest 
magnetic shear. However, contrary to the drawing of \cite{Melrose91}, the line of current 
reversal may not systematically correspond to that of the maximum velocity. For instance, 
the maximum photospheric velocity occurs at $|y| = y_0 = 1$ for the $\mathrm{S}\{1;2;3;4\}$ 
runs. Nevertheless, only $\mathrm{S}\{1;2\}$ possess a current reversal at $|y| \approx y_0$ 
at the photosphere, while it occurs at $|y| \lesssim 0.5$ for $\mathrm{S}3$ and at the PIL 
for $\mathrm{S}4$. This is because the current distribution depends not only on the shearing 
velocity profile but also on the 3D shape of magnetic field lines.

%:          Figure: Jz-sign shearing
%%%%%%%%%%%%%%%%%%%%%%%%%%%%%%%%%%%%%%%%%
%%%%%%%%%% FIGURE ?? %%%%%%%%%%%%%%%%%%%%%%%%%%
%%%%%%%%%%%%%%%%%%%%%%%%%%%%%%%%%%%%%%%%%
%%%%%%%%%%%%%%%%%%%%%%%%%%%%%%%%%%%%%%%%%
%%%%%%%%%%%%%%%%%%%%%%%%%%%%%%%%%%%%%%%%%

As for the twisting runs, we find that one significant difference between each shearing 
simulation is the extension of the currents close to the PIL. In particular, the shearing 
region of $\mathrm{S}1$ and $\mathrm{S}2$ is so narrow that their distribution 
of currents is strongly localized (in the $y$-direction) within the isocontour $| B_z | = 0.25$. 
On the contrary, the shearing region is so broad for $\mathrm{S}3$ and $\mathrm{S}4$ 
that the distribution of currents extends to the PIL. Finally, both the shearing profile 
and electric currents extend to the PIL for $\mathrm{SP}1$ and $\mathrm{SP}2$.

%%%%%%%%%%%%%%%%%%%%%%%%%%%%%%%%%%%%%%%%
%%%%%%%%%%%%%%%%%%%%%%%%%%%%%%%%%%%%%%%%

\subsection{Evolution of the total direct and return currents} \label{sec:S-Evolution-DRC-Shearing}

\begin{figure}
   \centering	 
  	 \centerline{\includegraphics[width=0.47\textwidth,clip=]{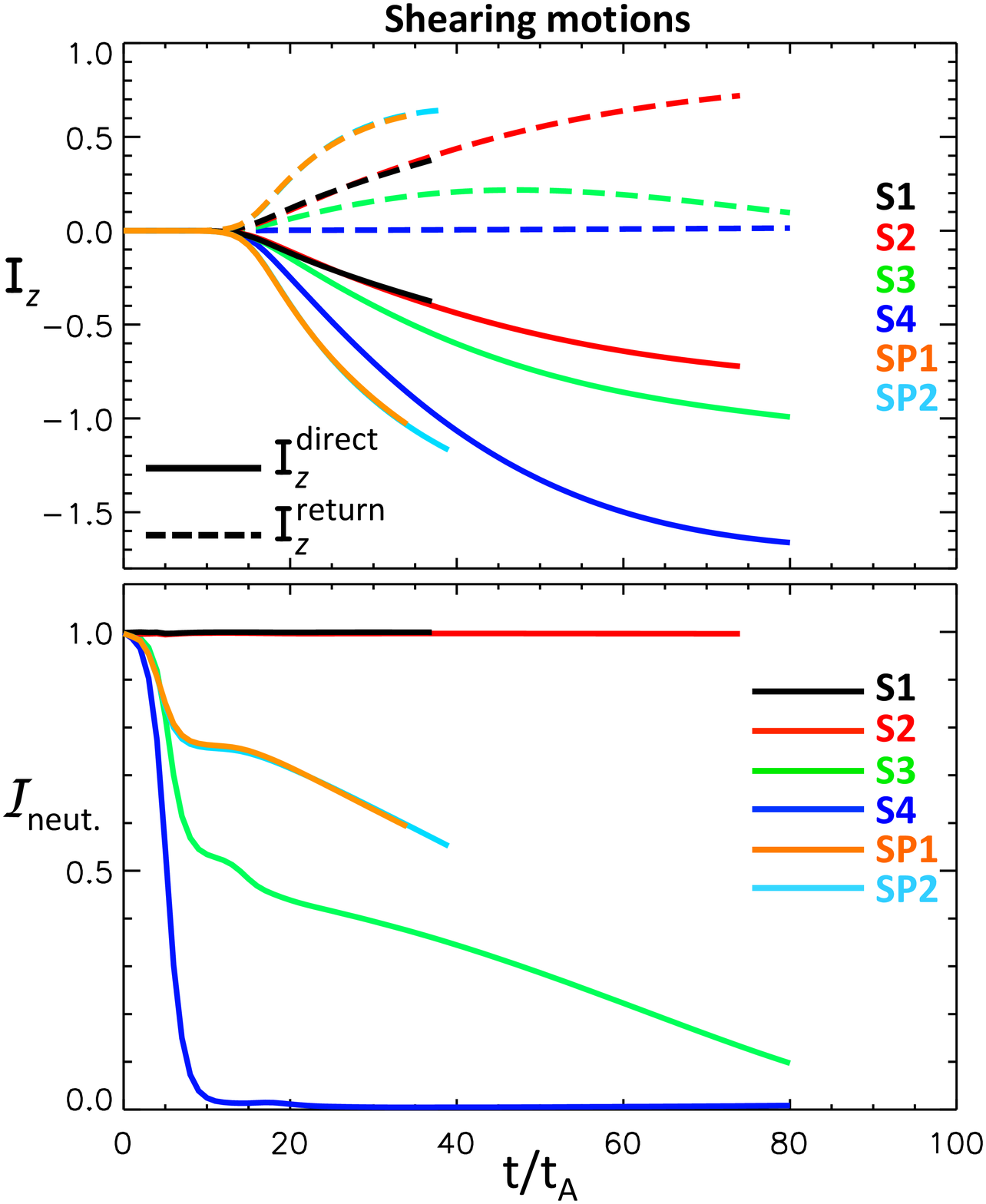}
	 }
   \caption{Evolution of electric current as in \fig{Fig-Iztwist} for the photospheric shearing motions.}
              \label{fig:Fig-Izshear}
    \end{figure}

The top panel of \fig{Fig-Izshear} presents the temporal evolution of the integrated 
direct and return currents in the positive magnetic polarity. All direct current curves 
show a similar monotonic rise in absolute value with a higher intensity for broader 
shearing regions. By contrast, the evolution of the return currents varies with the width 
of the shearing region. For $\mathrm{S}\{1;2\}$, the return current monotonically increases 
with time. For $\mathrm{S}{3}$, the intensity of the return current increases and then smoothly 
decreases, while for $\mathrm{SP}\{1;2\}$ the simulation terminated too early to draw 
conclusions. 
The $\mathrm{S}4$ run does not show any return current because the line of current 
reversal occurs at the PIL (\cf \sect{S-Shearing-JzDistribution}).

As shown in \app{A-RC-Decrease}, the applied shearing motions can generate 
two evolutionary phases for the current density of magnetic field lines: (1) a first 
phase of increase with magnetic shear, and (2) a phase of decrease caused 
by a fast elongation of the field lines. For similar shearing velocities, the shortest 
field lines should be the first to experience (2), because this effect is more pronounced 
for field lines that align more rapidly with the PIL (as explained in \app{A-RC-Decrease}). 
For all shearing simulations, the return currents are associated with the shortest magnetic 
field lines. We thus argue that it is the late elongation of these field lines that is responsible 
for the decrease of return current observed for $\mathrm{S}3$.

Finally, we note that increasing $\zeta_{sh2}$ for the second shearing profile 
(\ie $\mathrm{SP}2$), allows one to modify the shear profile of the field lines 
associated with the direct current. In particular, the direct current is distributed 
over a broader region. However, the amount of direct and return current is preserved.

%:          Figure: Iz & Neutralisation rate - shearing
%%%%%%%%%%%%%%%%%%%%%%%%%%%%%%%%%%%%%%%%%
%%%%%%%%%% FIGURE ?? %%%%%%%%%%%%%%%%%%%%%%%%%%
%%%%%%%%%%%%%%%%%%%%%%%%%%%%%%%%%%%%%%%%%
%%%%%%%%%%%%%%%%%%%%%%%%%%%%%%%%%%%%%%%%%
%%%%%%%%%%%%%%%%%%%%%%%%%%%%%%%%%%%%%%%%%

%%%%%%%%%%%%%%%%%%%%%%%%%%%%%%%%%%%%%%%%
%%%%%%%%%%%%%%%%%%%%%%%%%%%%%%%%%%%%%%%%

\subsection{Evolution of the neutralization ratio} \label{sec:S-Evolution-NRate-Shearing}

The bottom panel of \fig{Fig-Izshear} displays the temporal evolution of the neutralization 
ratio for each shearing run. For the reasons explained in \sect{S-Evolution-NRate-Twisting}, 
we focus our analysis on the neutralization during the constant photospheric driving 
phase (\ie $t \gtrsim 20 t_A$).

The currents remain fully neutralized ($\mathcal{I}_{\mathrm{neut.}} = 1$) for $\mathrm{S}1$ 
and $\mathrm{S}2$. 
On the contrary, the two runs with the broadest shearing widths ($\mathrm{S}\{3;4\}$) 
exhibit a strong departure from neutralization. For $\mathrm{S}4$, the neutralization 
ratio vanishes as a consequence of the absence of return current in the simulation 
(because the line of current reversal occurs at the PIL; \cf \sect{S-Shearing-JzDistribution}). 
For $\mathrm{SP}\{1;2\}$, the shearing motions applied close to the PIL also generate 
a strong net current. Both runs lead to the same degree of current neutralization.

We notice that the curves of neutralization ratio of the $\mathrm{S}3$ and $\mathrm{SP}\{1;2\}$ 
runs all present a comparable evolution. They all show a continuous decrease. Such 
a decrease indicates that the rate of current build-up is stronger for the direct current 
than for the return current. As previously mentioned (\sect{S-Evolution-DRC-Shearing}), 
the effect of field line elongation more strongly affects the shortest field lines. One then 
expects a lower rate of current build-up for the return current than for the direct one. 
This could then explain the continuous decrease of neutralization ratio observed 
for the $\mathrm{SP}\{1;2\}$ and $\mathrm{S}3$ runs.

Finally, the decreasing behavior of the neutralization ratio of $\mathrm{S}3$ and 
$\mathrm{SP}\{1;2\}$ indicates that, in a system driven by stationary shearing motions, 
the neutralization state of the system is not solely determined by the spatial properties 
of the shearing motions, but also depends on the amount of accumulated shear and 
inflation.

%%%%%%%%%%%%%%%%%%%%%%%%%%%%%%%%%%%%%%%%
%%%%%%%%%%%%%%%%%%%%%%%%%%%%%%%%%%%%%%%%
%%%%%%%%%%%%%%%%%%%%%%%%%%%%%%%%%%%%%%%%
%%%%%%%%%%%%%%%%%%%%%%%%%%%%%%%%%%%%%%%%

\section{Discussion} \label{sec:S-Discussion}

In this section, we first examine the development of weak compression currents 
in the ambient field of our simulations. We then discuss our results in the framework 
of the sheared-PIL/net current relationship and confront them to the conjectures 
of \cite{Melrose91} and \cite{Parker96}.

%%%%%%%%%%%%%%%%%%%%%%%%%%%%%%%%%%%%%%%%
%%%%%%%%%%%%%%%%%%%%%%%%%%%%%%%%%%%%%%%%

\subsection{Compression currents in the ambient field} \label{sec:S-Compression-Currents}

\begin{figure}
   \centering
  	 \centerline{\includegraphics[width=0.47\textwidth,clip=]{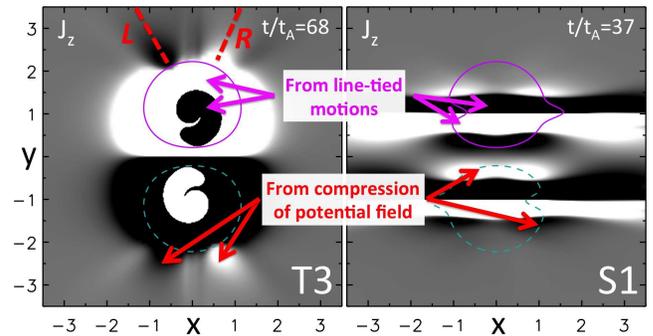}
	 } 	 	 
   \caption{Saturated photospheric $j_z$ (gray shading) showing the currents associated with the compression of the ambient field caused by the inflation of the twisted (left) and sheared (right) magnetic fields. The saturation value is $2.4 \times 10^{-3}$ for $\mathrm{T}3$ and $5 \times 10^{-3}$ for $\mathrm{S}1$ (compared with \figs{Fig-Jztwist}{Fig-Jzshear} where the saturation value is two orders of magnitude higher). The solid-purple and cyan-dashed lines are $|B_z| = 0.25$-isocontours.}
              \label{fig:Fig-3D-compression-currents}
    \end{figure}

When analyzing the distribution of currents of our 3D current-carrying fields at high 
saturation levels, we find that some very weak currents develop in the ambient field 
rooted in areas where the photospheric flows vanish (as pointed by the red arrows 
in \fig{Fig-3D-compression-currents}). These currents are typically $2-3$ orders 
of magnitude smaller than the currents directly generated by the photospheric flows. 
They appear in the very close vicinity of the twisted/sheared fields and very rapidly 
decrease away from them. We find that these currents are induced by the local 
compression of the ambient field caused by the inflation of the twisted/sheared field 
in response to the photospheric flows. Such a compression of the ambient field cannot 
be reproduced in 2.5D geometry because the imposed symmetry forces that field to stay 
potential.

%:          Figure: Currents from 3D compression of Bp
%%%%%%%%%%%%%%%%%%%%%%%%%%%%%%%%%%%%%%%%%
%%%%%%%%%% FIGURE ?? %%%%%%%%%%%%%%%%%%%%%%%%%%	
%%%%%%%%%%%%%%%%%%%%%%%%%%%%%%%%%%%%%%%%%
%%%%%%%%%%%%%%%%%%%%%%%%%%%%%%%%%%%%%%%%%
%%%%%%%%%%%%%%%%%%%%%%%%%%%%%%%%%%%%%%%%%

We also note that the compression currents do not form a shell of a single sign around 
the return current for the twisting cases (\cf $\mathrm{T}3$ in \fig{Fig-3D-compression-currents}). 
On the contrary, we find two regions of enhanced currents oriented in two specific directions 
(as indicated by the dashed red lines in \fig{Fig-3D-compression-currents}). This is caused 
by the development of a kink of the flux tube axis. The direction of the $L$-dashed red line 
corresponds to the orientation of the kink of the flux tube axis 
\citep[as shown in \fig{Fig-B3D} top left and more precisely by the yellow field line in Figure 5 of][]{Aulanier05}. 
The kink of the axis causes a preferential compression of the ambient field in its direction, 
and induces a magnetic depression in the direction of the $R$-dashed red line.	

%%%%%%%%%%%%%%%%%%%%%%%%%%%%%%%%%%%%%%%%
%%%%%%%%%%%%%%%%%%%%%%%%%%%%%%%%%%%%%%%%

\subsection{The sheared-PIL/net current relationship} \label{sec:S-ShearPIL-NetI}

\begin{figure}
   \centering
  	 \centerline{\includegraphics[width=0.47\textwidth,clip=]{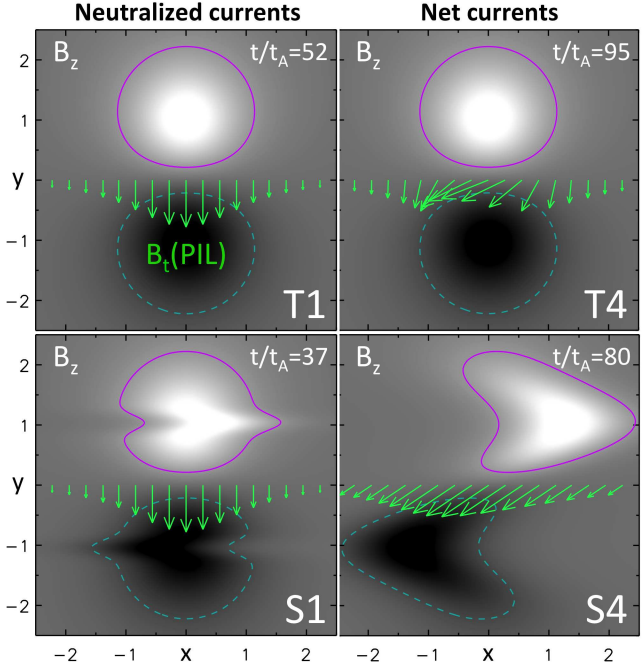}
	 } 	 	 
   \caption{Sheared-PIL/net current relationship illustrated for two current-neutralized ($\mathrm{T}1$ and $\mathrm{S}1$; left column) and two net current ($\mathrm{T}4$ and $\mathrm{S}4$; right column) simulations. The photospheric transverse magnetic field at the PIL, $\bb_t (\mathrm{PIL})$ (green arrows), is plotted over the photospheric $B_z$ (gray shading). The solid-purple and cyan-dashed lines are $|B_z| = 0.25$-isocontours.}
              \label{fig:Fig-Neutralized-vs-Net-Currents}
    \end{figure}

The numerical experiments of \sects{S-Twisting-Results}{S-Shearing-Results} show 
that photospheric line-tied motions can generate 3D force-free magnetic fields with 
different amount of current neutralization. We further analyzed and compared 
the simulations presented in this paper to identify the origin of these various degrees 
of neutralization. The results are summarized in \fig{Fig-Neutralized-vs-Net-Currents} 
that displays the photospheric transverse magnetic field at the PIL for current-neutralized 
and net current cases. We find that all current-neutralized magnetic fields possess a PIL 
fully embedded in a potential magnetic field (\fig{Fig-Neutralized-vs-Net-Currents} left 
column). On the contrary, we observe that (1) a net current develops simultaneously 
with magnetic shear at the PIL, and (2) a stronger net current is induced for simulations 
generating a stronger magnetic shear at the PIL (see \fig{Fig-Neutralized-vs-Net-Currents} 
right column). In agreement with \cite{Torok03}, and extending their results to pure 
shearing motions, we thus find that a net current solely develops when magnetic shear 
is built up at the PIL. These results are also consistent with observational studies 
\citep[\eg][]{Ravindra11,Georgoulis12} and the investigation of flux emergence in 
\cite{Torok14}. In the following, we address this relationship in a general form.

%:          Figure: Neutralized vs Net Currents
%%%%%%%%%%%%%%%%%%%%%%%%%%%%%%%%%%%%%%%%%
%%%%%%%%%% FIGURE ?? %%%%%%%%%%%%%%%%%%%%%%%%%%	
%%%%%%%%%%%%%%%%%%%%%%%%%%%%%%%%%%%%%%%%%
%%%%%%%%%%%%%%%%%%%%%%%%%%%%%%%%%%%%%%%%%
%%%%%%%%%%%%%%%%%%%%%%%%%%%%%%%%%%%%%%%%%	

\begin{figure}
   \centering
  	 \centerline{\includegraphics[width=0.4\textwidth,clip=]{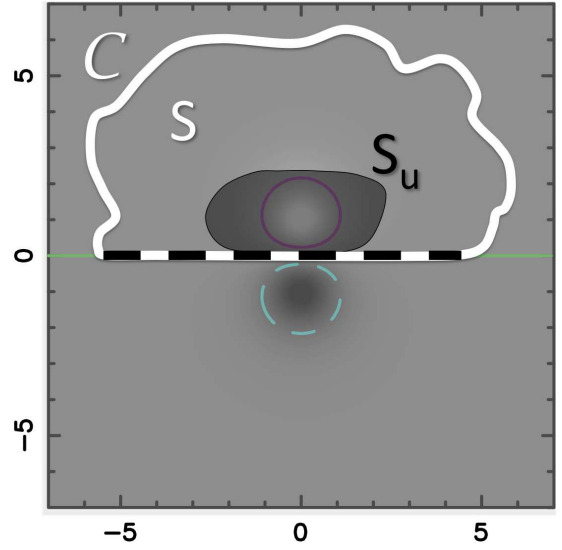}
	 } 
   \caption{Schematic used for the mathematical demonstration of the sheared-PIL/net current relationship. The map shows a photospheric magnetogram of the initial potential magnetic field used in our line-tied simulations. The surfaces $S/S_u$ and the contour $\mathcal{C}$ are defined in \sect{S-ShearPIL-NetI}. The thick, black/white, dashed line highlights the part of $\mathcal{C}$ that corresponds to the PIL (green solid line). The solid purple, and dashed cyan, lines are $B_z = \pm 0.25$-isocontours of the magnetic field.}
              \label{fig:Fig-ShearedPIL-NetI}
    \end{figure}

We consider a bipolar potential magnetic field such as the initial field used in 
our numerical simulations (\fig{Fig-ShearedPIL-NetI}). We then consider 
a photospheric velocity field, $\uu$, that builds up electric currents at $z=0$ inside 
of a surface, $S_{u}$, of the positive magnetic polarity\footnote{The corresponding 
definition can also be done in the negative magnetic polarity, but this is not needed 
because of $\grad \cdot \jj =0$ and current is transferred to closed magnetic field lines. 
We therefore keep our analysis simpler by focusing on the positive magnetic polarity, 
as we did with the analysis of our numerical simulations.}; $\uu$ vanishes outside of 
$S_{u}$. Finally, we consider a closed curve, $\mathcal{C}$, such that (i) $\mathcal{C}$ 
includes the PIL of the bipolar AR, and (ii) the surface, $S$, enclosed by $\mathcal{C}$ is 
much larger than the surface $S_{u}$ where electric currents are generated by the photospheric 
motions. These two choices ensure that all the currents transferred to the magnetic field are 
fully enclosed by $\mathcal{C}$.

The applied photospheric velocity field generates a current-carrying component, 
$\bb_{j}$, in the bipolar magnetic field. The total magnetic field, $\bb$, can be 
decomposed as the sum of its potential, $\bbp$, and current-carrying, $\bb_{j}$, 
components \citep[\eg][]{Valori13}, such that
	\BA 		\label{eq:Eq-B-Decomposition}
     		\bb & = & \bbp + \bb_{j}   \,,     \\  
		\curl \bb & = & \curl \bb_j = \mu_0 \jj   \,.
		 	\label{eq:Eq-J-Decomposition}
     	\EA
In fact, $\bb_j$ can be further decomposed into two components
	\BE 		\label{eq:Eq-Bj-Decomposition}
     		\bb_j = \bb^{\uu \ne 0}_j + \bb^{\uu = 0}_{j}   \,.
     	\EE
$\bb^{\uu \ne 0}_j$ corresponds to the main current-carrying field and it has non-zero 
values only in the magnetic field connected to $S_u$. $\bb^{\uu = 0}_j$ is a current-carrying 
field generated by the compression of the potential field resulting from the inflation of 
the main current-carrying field. This component is non-zero only in the ambient magnetic 
field, \ie the magnetic field not connected to $S_u$. Its strength is on average two orders 
of magnitude smaller than the strength of the main current-carrying field, $\bb^{\uu \ne 0}_j$, 
and decreases very rapidly away from the latter (\cf \sect{S-Compression-Currents}). 
For this reason and because the contour $\mathcal{C}$ is chosen such that $S$ is much 
larger than $S_{u}$, we can neglect $\bb^{\uu = 0}_j$ in the following.

%:          Figure: Sheared-PIL / Net Current Scheme
%%%%%%%%%%%%%%%%%%%%%%%%%%%%%%%%%%%%%%%%%
%%%%%%%%%% FIGURE ?? %%%%%%%%%%%%%%%%%%%%%%%%%%	
%%%%%%%%%%%%%%%%%%%%%%%%%%%%%%%%%%%%%%%%%
%%%%%%%%%%%%%%%%%%%%%%%%%%%%%%%%%%%%%%%%%
%%%%%%%%%%%%%%%%%%%%%%%%%%%%%%%%%%%%%%%%%

Applying Amp\`ere's law, it then follows that the total electric current enclosed by 
$\mathcal{C}$ is
	\BA
		I & = & \frac{1}{\mu_0}  \oint_{\mathcal{C}}   \bb_{j}  \cdot \dl  \, \nonumber  \\
		 & \approx & \frac{1}{\mu_0}  \oint_{\mathcal{C}}   \bb^{\uu \ne 0}_{j}  \cdot \dl   \,,
		             \label{eq:Eq-I-Shear-R-Init}
     	\EA
where $\dl$ is the line element along $\mathcal{C}$. \eq{Eq-I-Shear-R-Init} 
can be further decomposed as the sum of two contributions: one along the part of 
$\mathcal{C}$ corresponding to the PIL, and a second corresponding to the remaining 
of $\mathcal{C}$ (\ie $\mathcal{C}-PIL$). This leads
	\BE
     		I \approx \frac{1}{\mu_0} \left( \int_{PIL}   \bb^{\uu \ne 0}_j  \cdot \dl \ + \  \int_{\mathcal{C}-PIL}  \bb^{\uu \ne 0}_j  \cdot \dl \right)  \,.
		             \label{eq:Eq-I-Shear-R-TMP}
     	\EE
The intersection of $\mathcal{C}$ with $S_{u}$ --- when it exists --- is limited to the PIL, 
and $\bb^{\uu \ne 0}_j$ vanishes outside $S_{u}$. It then follows that the second term 
of \eq{Eq-I-Shear-R-TMP} vanishes. At any fixed time, the total current enclosed 
by $\mathcal{C}$ is therefore
	\BE
     		I \approx \frac{1}{\mu_0} \int_{PIL}   \bb^{\uu \ne 0}_j  \cdot \mathbf{dl}   \,.
		             \label{eq:Eq-I-Shear-R}
     	\EE
When the PIL is embedded within a potential magnetic field, \ie $S_u$ is never in contact 
with the PIL, then $\bb^{\uu \ne 0}_j$ vanishes all along the PIL. The total electric current, 
$I$, thus vanishes as well, which implies that the currents are neutralized. On the contrary, 
$\bb^{\uu \ne 0}_j$ does not vanish for a magnetically-sheared PIL. The current-carrying 
magnetic field will then contain a net electric current, $I \ne 0$ (unless the integrand 
in \eq{Eq-I-Shear-R} changes sign such that the oppositely directed contributions cancel 
exactly, which could happen for the very rare cases of ARs containing opposite twist/shear). 
We thus conclude that {\it force-free} net currents are inevitably related to magnetically 
sheared PILs.

The above derivation is supported by the results of the $\mathrm{SP2}$ run as compared 
with $\mathrm{SP}1$. Indeed, both possess the same shearing profile in the field lines 
associated with the return current. This means that both simulations have approximately 
the same magnetic shear profile along the PIL (neglecting the effect of different distant direct 
currents). From \eq{Eq-I-Shear-R}, it follows that both are expected to have very similar 
net current. This is what happens as inferred in \fig{Fig-Izshear} from the curves of direct 
current, return current, and neutralization ratio of $\mathrm{SP}2$ which match those of 
$\mathrm{SP}1$.

Finally, we emphasize that the above relationship between net currents and magnetically 
sheared PIL is significantly different from the Lorentz-force-driven shear discussed 
by \cite{Georgoulis12}. Indeed, in all our zero-$\beta$ MHD simulations, there is 
no dynamical compression that could generate a Lorentz force which would shear the PIL. 
On the contrary, the magnetic shear along the PIL is caused by the motions imposed 
in the photosphere, and this magnetic shear generates a {\it force-free} net current.

%%%%%%%%%%%%%%%%%%%%%%%%%%%%%%%%%%%%%%%%
%%%%%%%%%%%%%%%%%%%%%%%%%%%%%%%%%%%%%%%%

\subsection{Net currents versus neutralization predictions} \label{sec:S-NetI-vs-NPredictions}

The possibility of generating coronal magnetic fields carrying a net current is at variance 
with the conjectures of \cite{Melrose91} and \cite{Parker96} who argued that twisted 
and sheared coronal fields should be perfectly current-neutralized. The main difference 
with our results relies in the fact that both authors built a conjecture using 2.5D 
considerations which were then directly transposed to 3D geometry. The cornerstone 
of their conjectures is that any 3D twisted/sheared magnetic field embedded in a potential 
field or field-free environment can be equivalently described by a flux tube connecting 
two parallel planes and set in the same (but 2.5D) magnetic environment.

The above assumption is valid for any 3D flux tube whose two photospheric magnetic 
polarities are fully separated either by an ambient potential field or a field-free environment, 
\ie when there is no magnetic shear at the PIL. This is the implicit assumption of 
\cite{Melrose91} who limited his considerations to the case of twisting and shearing 
motions not extending to the PIL. The same assumption is also implicitly made 
by \cite{Parker96} when using a cylindrical twisted flux tube of finite radius as a model 
for an element of an AR magnetic field which is fragmented in the photosphere. In such cases, 
a 2.5D analysis shows that electric currents should be neutralized (see \app{A-Currents-2.5D}). 
This agrees with our 3D derivation of \sect{S-ShearPIL-NetI} in which we demonstrated 
that full current neutralization occurs when a PIL is fully embedded in a potential field or 
field-free environment.

On the contrary, when a 3D flux tube has magnetic shear along its PIL, the above 
cylindrical description of an AR is not valid. Indeed, an electric current is then flowing 
along the PIL where the two polarities are in contact. Then, the cylindrical approximation 
of \cite{Melrose91} and \cite{Parker96} is not relevant. As a consequence, the associated 
conclusion that currents should be neutralized is not applicable. A fully 3D analysis is 
then required to predict whether or not currents should be neutralized. Such a 3D analysis 
actually show that a net current should be expected when magnetic shear is present along 
the PIL of the 3D flux tube (\cf \sect{S-ShearPIL-NetI}).

%%%%%%%%%%%%%%%%%%%%%%%%%%%%%%%%%%%%%%%%
%%%%%%%%%%%%%%%%%%%%%%%%%%%%%%%%%%%%%%%%
%%%%%%%%%%%%%%%%%%%%%%%%%%%%%%%%%%%%%%%%
%%%%%%%%%%%%%%%%%%%%%%%%%%%%%%%%%%%%%%%%

\section{Conclusions} \label{sec:S-Conclusions}

%
% Brief summary
%
In this study, we used 3D MHD numerical simulations to analyze the properties 
of electric currents in line-tied coronal fields generated by photospheric flows in bipolar 
ARs. We showed that typical photospheric flows, such as twisting and shearing motions, 
invariably produce both {\it direct} and {\it return} currents in the line-tied coronal fields. 
We find that these photospheric flows can create both neutralized and non-neutralized 
currents.

%
% Origin of net currents
%
Using Amp\`ere's law, we provided a physical origin to the build-up of {\it force-free} 
net currents in coronal magnetic fields. They arise from the development of magnetic 
shear along PILs (\sect{S-ShearPIL-NetI}). This conclusion agrees with the independent 
study of \cite{Torok14} who showed that magnetic flux emergence can also produce 
a net coronal current which simultaneously develops with magnetic shear at the PIL. 
\cite{Torok14} and our study thus set a theoretical framework for understanding 
the properties of electric currents in ARs. They both show that, in general, net currents 
can be formed in the corona from various, independent and/or combined processes: 
\eg magnetic flux emergence, photospheric flows, and by extension, any mechanism 
that can generate magnetic shear along a PIL.

%
% Predictions for observations
%
Net currents in ARs can therefore develop in a large variety of cases. On the contrary, 
the production of perfectly current-neutralized magnetic fields in the solar atmosphere 
is a special case that requires a rather uncommon condition: a main PIL without magnetic 
shear. This is unlikely for both emerging and evolved/decaying ARs. Indeed, several 
observational and numerical studies show that newly emerged ARs generally possess 
a strongly sheared PIL \citep[\eg][]{Manchester04,Canou09,Georgoulis12,Toriumi13,Poisson15}. 
Next, evolved/decaying ARs are often associated with the presence of H$\alpha$ filaments, 
which are cold dense material supported in highly sheared/twisted magnetic fields lying 
above PILs \citep[\eg][]{VanBallegooijen89,Aulanier98,Gibson04,Schmieder06,Jing10,Mackay10,Xia14}. 

Furthermore, starting from a configuration with an un-sheared PIL, it will remain so if 
the shear component of photospheric flows is not extending to the PIL. This is also unlikely 
since the opposite is typically observed in ARs \citep[\eg][]{Vemareddy12,Wang12,Guo13,Liu13}. 
In fact, several observations suggest shearing profiles that would be analogous to our strongly 
non-neutralized run $\mathrm{S}4$ 
\citep[as inferred from the photospheric motions of magnetic polarities; \eg][]{Su07,Sun12}. 
Finally, the absence of significant shear is more likely to hold for evolved ARs possessing 
isolated magnetic polarities (\ie magnetic polarities far away from each other).

%
% Real neutralization VS weak net currents
%
The two main sources of AR currents, emergence and horizontal photospheric motions, 
would thus be expected to primarily produce net coronal currents. Nevertheless, a few 
observational studies seem to indicate that ARs with neutralized currents may be as numerous 
as ARs with a net current \cite[\eg][]{Wilkinson92,Wheatland00}. If we consider a typical 
magnetic field of $B_{\mathrm{max}} = 2000 \ \mathrm{G}$ for newly emerged ARs, and 
$15 \ \mathrm{Mm}$ for one spatial unit of our non-dimensionalized simulations, the strength 
of the net current in our strongly non-neutralized magnetic fields can reach 
$\sim 0.7 - 4 \ \times 10^{12} \ \mathrm{A}$. If the magnetic field is scaled to 
$B_{\mathrm{max}} = 200 \ \mathrm{G}$ for decaying ARs, the strength of the net current 
ranges between $\sim 0.7 - 4 \ \times 10^{11} \ \mathrm{A}$. Since the actual measurement 
precision is about $10^{11} \ \mathrm{A}$, it then remains to be proved that current-neutralized 
ARs are truly current-neutralized. This requires a systematic analysis of both the current-neutralization 
and the magnetic shear at the PIL for any studied AR, which was not done in \cite{Wilkinson92} 
and \cite{Wheatland00}. Such dedicated studies could then be used to further test the conclusions 
derived in the present paper.

%
% Net currents and CME modeling
%
Finally, even though our MHD simulations systematically report the presence of return 
currents, net currents in coronal magnetic fields do exist. Therefore, eruption models 
based on magnetic configurations possessing net currents 
\citep[\eg][]{VanTend78,Heyvaerts82,Lin98,Titov99,Kliem06,Demoulin10} 
are a simplified, but valid, description of pre-eruptive magnetic fields in ARs. Our results 
that relate net coronal currents to magnetically-sheared PILs then naturally explain 
the observational conclusions of \cite{Falconer01} and \cite{Falconer02}, which state 
that ARs with a sheared PIL are more CME-prolific. That being said, one must bear in mind 
that the aforementioned analytical models, based on a net coronal current, do not possess 
any return current. Yet, return currents exist in MHD simulations of CMEs 
\citep[\eg][]{Torok03,Delannee08,Aulanier12}. Moreover, in some cases the return current 
can have the same strength as the direct current, which may inhibit the eruption \cite[\eg][]{Forbes10}. 
Further studies are then required to quantify the role of these return currents for the trigger 
and development of solar flares and CMEs, \eg using MHD simulations.

\begin{acknowledgements}
We thank the anonymous referee for a careful consideration of the manuscript 
and helpful comments. We thank V. S. Titov for insightful discussions. 
The calculations presented in this work were performed on the quadri-core bi-Xeon 
computers of the Cluster of the Division Informatique de l'Observatoire de Paris. 
K.D. acknowledges funding from the Computational and Information Systems 
Laboratory and from the High Altitude Observatory, as well as support from 
the Air Force Office of Scientific Research under award FA9550-15-1-0030. 
The National Center for Atmospheric Research is sponsored by the National 
Science Foundation. T.T. was supported by NSF through grant AGS-1348577 
and by NASA's LWS program. B.K. acknowledges support by the DFG.
\end{acknowledgements}

%%% BIBLIOGRAPHY %%%%%%%%%%%%%%%%%%%%%%%%%%%%%%%%%%%%%%%%%%%%%%%%%%%%%%%%%%%	
	    % format of references provided by the journal (.bst)
\bibliographystyle{aa}
      
     % name your Bibtex file containing your references (.bib)
\bibliography{ReturnCurrents}  

     % Checking: look if the file containing the ``\bibitem'' exits
     %           so check if the .bbl file exist (bibTeX compilation)
\IfFileExists{\jobname.bbl}{} {\typeout{}
\typeout{****************************************************}
\typeout{****************************************************}
\typeout{** Please run "bibtex \jobname" to obtain} \typeout{**
the bibliography and then re-run LaTeX} \typeout{** twice to fix
the references !}
\typeout{****************************************************}
\typeout{****************************************************}
\typeout{}}

%%% APPENDICES %%%%%%%%%%%%%%%%%%%%%%%%%%%%%%%%%%%%%%%%%%%%%%%%%%%%%%%%%%%	
\appendix

\section{A. Electric currents in cylindrical geometry} \label{app:A-Currents-2.5D}

Let us consider a twisted magnetic flux tube in cylindrical geometry. The cylindrical 
coordinates $r$, $\theta$, and $z$, respectively describe the distance to the axis of 
the flux tube, the angle around its axis, and the position along its axis. Using the integrated 
form of Amp\`ere's law, the total current, $I$, flowing across a disk of radius $r$, is
	\BE   \label{eq:Eq-Current-2.5D}
     		\mu_0 \ f\ I (r) = 2 \pi \ r \ B_{\theta} (r) \,,
	\EE
where $f$ (equals $+1$ or $-1$) is included so that $I (r)$ is a positive function in the flux 
rope core where the direct current is located. In the flux rope core, the current $I$ is 
therefore a growing function of $r$. The region of return current is located where $I$ is 
a decreasing function of $r$, so for
 	\BE   \label{eq:Eq-retcurr-condition-TMP}
		\pder{ \left(r B_{\theta} \right) }{r} < 0 \,.
	\EE
The existence of return currents is thus simply constrained by the existence of $r_l>0$, 
such that
	\BE   \label{eq:Eq-retcurr-condition}
     		B_{\theta} (r) < \frac{C}{r} \,, \ \mathrm{for} \ 
		r > r_l \,,
     	\EE
where $C$ is an integration constant. The region where $B_{\theta}$ decreases faster 
than $1/r$ is the region of return current.

The flux rope current is fully neutralized if there is a finite radius, $r_c$, such that 
$I( r > r_c ) = 0$, \ie
	\BE   \label{eq:Eq-neutralization-condition}
     		B_{\theta} (r) = 0 \,, \ \mathrm{for} \ 
		r > r_c \,.
     	\EE
From \eq{Eq-Current-2.5D}, it is straightforward to show that if a twisted magnetic flux 
tube is confined (\ie if it has a finite radius $R=r_c$), then the total electric current carried 
by the flux tube vanishes.

These conditions are valid regardless of the force-freeness of the magnetic field.

%%%%%%%%%%%%%%%%%%%%%%%%%%%%%%%%%%%%%%%%
%%%%%%%%%%%%%%%%%%%%%%%%%%%%%%%%%%%%%%%%
%%%%%%%%%%%%%%%%%%%%%%%%%%%%%%%%%%%%%%%%
%%%%%%%%%%%%%%%%%%%%%%%%%%%%%%%%%%%%%%%%

\section{B. Solenoidal conditions with OHM} \label{app:A-Solenoidal-Condition}

%:          Figure: PDF solenoidal condition
%%%%%%%%%%%%%%%%%%%%%%%%%%%%%%%%%%%%%%%%%
%%%%%%%%%% FIGURE 11 %%%%%%%%%%%%%%%%%%%%%%%%%%	
   \begin{figure}
   \centering
  	 \centerline{\includegraphics[width=0.7\textwidth,clip=]{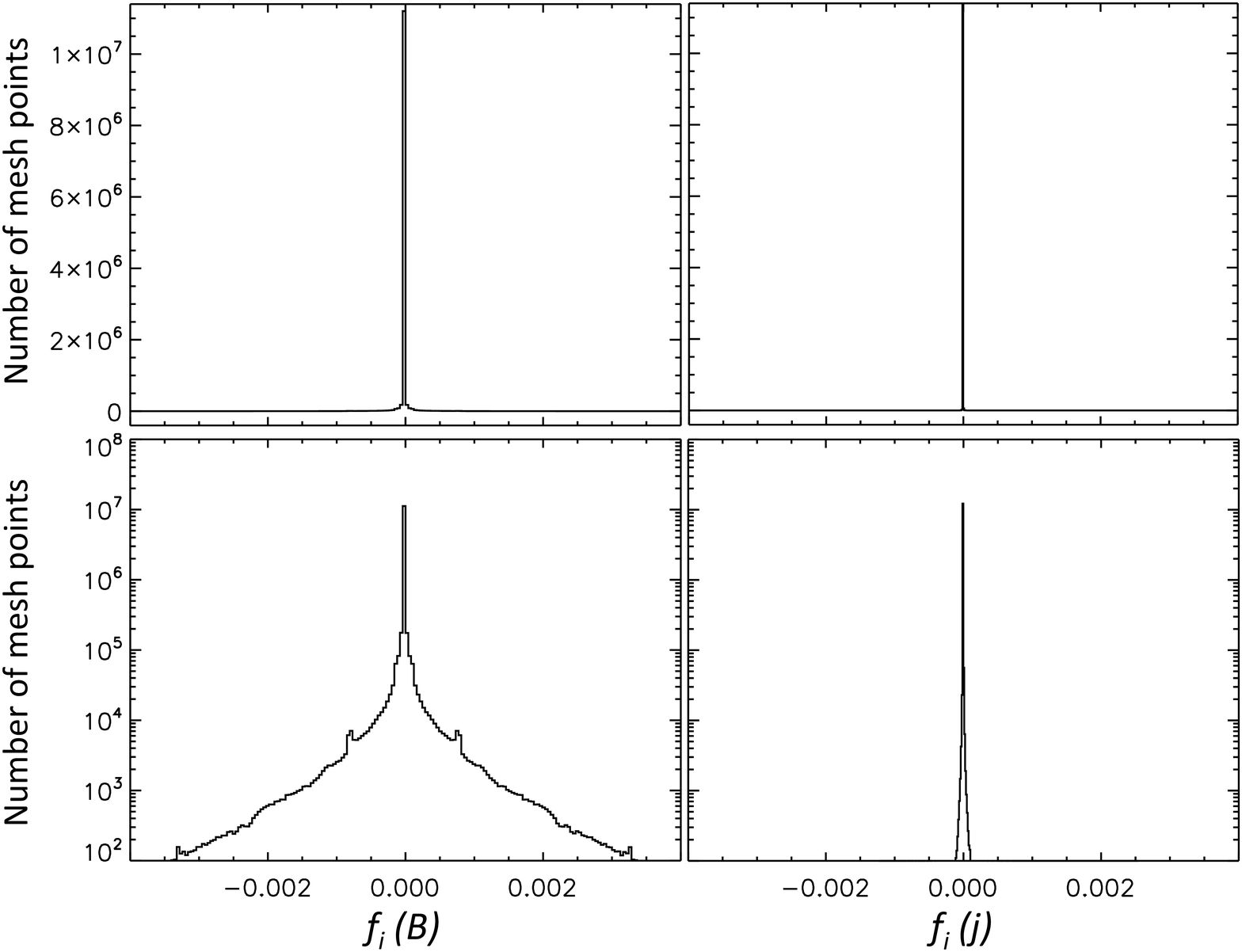}
	 } 
   \caption{PDF of the fractional flux, $f_i$ (\eq{Eq-fractional-flux}), of the magnetic field (left) and of the electric current density (right), for the twisting simulation $\mathrm{T}4$ at $t=60 \ t_A$. Top: linear scale. Bottom: log-scale.}
              \label{fig:Fig-PDF-Solenoidal-Conditions}
    \end{figure}
%%%%%%%%%%%%%%%%%%%%%%%%%%%%%%%%%%%%%%%%%
%%%%%%%%%%%%%%%%%%%%%%%%%%%%%%%%%%%%%%%%%
%%%%%%%%%%%%%%%%%%%%%%%%%%%%%%%%%%%%%%%%%

As mentioned in \sect{S-MHD-eq}, the solenoidal condition for the magnetic field is 
not numerically imposed in the code. However, to show that it does not affect the evolution 
of the magnetic field in our simulations, we compute the fraction of non-conserved flux, 
or fractional flux, $f_i$, within each cell, $i$, of the mesh, such that
	\BA
     		f_i (\vec{X}) & = & \frac{\int_{v_i}  \nabla \cdot \vec{X}_i  \rmd v_i}{ \int_{s_i}  | \vec{X}_i |  \rmd s_i}     \,   \\
     		 & \approx & \frac{v_i}{s_i} \frac{\nabla \cdot \vec{X}_i}{| \vec{X}_i |} \,,
			\label{eq:Eq-fractional-flux}
     	\EA
where $| \vec{X}_i |$ and $\nabla \cdot \vec{X}_i$ are respectively the norm of $\vec{X}$, 
and its divergence, within the mesh-cell, $i$, of volume, $v_i$, bounded by the surface, 
$s_i$ \citep[\eg][]{Wheatland00b,Valori13}.

We compute the set of $f_i (\bb)$ values for the twisted flux tube simulation, 
$\mathrm{T}4$. The corresponding probability density function (PDF) and its statistics 
are displayed in \fig{Fig-PDF-Solenoidal-Conditions} and \tab{Tab-Fractional-Flux-Metrics}. 
As one can see, the PDF is well centered on zero. The two bumps in the wings of the PDF 
are contributions from the boundaries.

%:          Table: Statistics of the PDF of fractional flux
%%%%%%%%%%%%%%%%%%%%%%%%%%%%%%%%%%%%%%%%%
%%%%%%%%%%%%%%%%%% Table ?? %%%%%%%%%%%%%%%%%
   \begin{deluxetable}{c c c c c c}
   \tablecaption{Statistics of the pdf of fractional flux, $f_i$.
   	\label{tab:Tab-Fractional-Flux-Metrics}
	}
   \tablehead{\colhead{$\vec{X}$} & \colhead{$<f_i (\vec{X})>$} & \colhead{median of $|f_i (\vec{X})|$} & \colhead{$< | f_i (\vec{X}) | >$} & \colhead{$\sigma (f_i (\vec{X}))$} & \colhead{$\sqrt{ < f(\vec{X})^{2}_{i} > }$}}
   \tablecomments{$\bb$ and $\jj$ respectively are the magnetic field and the electric current density. $<>$ and $\sigma$ respectively are the mean and standard deviation. Results are presented for the simulation $\mathrm{T}4$ at $t=60 \ t_A$.}
   \startdata
	$\bb$ & $-1 \times 10^{-10}$ & $2.3 \times 10^{-7}$ & $2.8 \times 10^{-5}$ & $1.7 \times 10^{-4}$ & $1.7 \times 10^{-4}$   \\
	$\jj$ & $2.1 \times 10^{-11}$ & $3.3 \times 10^{-8}$ & $2.9 \times 10^{-7}$ & $2.5 \times 10^{-6}$ & $2.5 \times 10^{-6}$
   \enddata
   \end{deluxetable} 
%%%%%%%%%%%%%%%%%%%%%%%%%%%%%%%%%%%%%%%%%
%%%%%%%%%%%%%%%%%%%%%%%%%%%%%%%%%%%%%%%%%

Both \fig{Fig-PDF-Solenoidal-Conditions} and \tab{Tab-Fractional-Flux-Metrics} show 
a mean, a median, and a standard deviation of $f_i$, that are all smaller than $\sim 10^{-4}$. 
This means that the amount of non-conserved magnetic flux within each mesh-cell is 
typically $\sim 10^{4}$ times weaker than the local magnetic flux. Therefore, even though 
the solenoidal condition is not numerically treated within the OHM code, the non-conservation 
of magnetic flux remains very weak. These results are representative of the set of numerical 
simulations performed and presented in this paper. We thus conclude that the solenoidal 
condition is well enough verified to ensure that the very weak non-conservation of magnetic 
flux does not affect the evolution of the system in each of our line-tied simulations.

We further compute the fractional flux for the electric current density ($f_i (\jj)$) to check 
that its divergence is indeed zero (as one would expect from applying the divergence 
operator to Amp\`ere's law). The associated PDF, and the statistics of the PDF, are 
displayed in \fig{Fig-PDF-Solenoidal-Conditions} and \tab{Tab-Fractional-Flux-Metrics}. 
As for the magnetic field, $\nabla \cdot \jj = 0$ is well preserved in our line-tied simulations.

%%%%%%%%%%%%%%%%%%%%%%%%%%%%%%%%%%%%%%%%
%%%%%%%%%%%%%%%%%%%%%%%%%%%%%%%%%%%%%%%%
%%%%%%%%%%%%%%%%%%%%%%%%%%%%%%%%%%%%%%%%
%%%%%%%%%%%%%%%%%%%%%%%%%%%%%%%%%%%%%%%%

\section{C. Neutralization ratio: transition phase} \label{app:A-Nratio-Transition-phase}

The neutralization ratio curves of the partially current-neutralized cases all present 
a transition phase between $t = 0$ and $t \approx 20 t_A$. This transition phase is 
distinguished by two specific periods, $t = [0;10] t_A$ and $t = [10;20] t_A$.

As mentioned \sect{S-Evolution-DRC-Twisting}, the direct/return currents are computed 
by extracting the negative/positive current density at the photospheric positive polarity. 
During the early evolution of the system (when the driving is zero and/or extremely weak), 
the computed total direct/return current, and hence neutralization ratio, are all dominated 
by the noise. This noise is essentially due to the presence of magneto-acoustic waves 
at early times. While the initial potential field is analytically potential and at equilibrium, 
it is not numerically because of the discretization. This well-known effect leads to 
the generation of magneto-acoustic waves inducing compression of the magnetic field. 
Such a compression creates transitory neutralized currents in each magnetic polarity 
of the system because they are not associated with magnetic shear build-up at the PIL. 
This is why the neutralization ratio is initially well-defined and equal to 1 for all simulations. 
A precise analysis shows us that the currents generated by the photospheric motions 
progressively become dominant (\ie their strength is $\sim 100$ times larger than the noise) 
for $t \gtrsim 10 t_A$. The transition from a system dominated by noise currents towards 
photospherically-generated currents is thus responsible for the early evolution of 
the neutralization ratio, up to $10 t_A$.

The second transition appears between $t = [10;20] t_A$, which corresponds 
to the main acceleration period of the temporal ramp function. This transition follows 
the evolution of the temporal ramp function. Such a transition could be produced by 
two competitive mechanisms: (1) non-force-free effects due to the fast acceleration 
of the photospheric velocities, and (2) the saturation of currents due to field line length 
(as discussed in \sects{S-Evolution-DRC-Twisting}{S-Evolution-DRC-Shearing}).

%%%%%%%%%%%%%%%%%%%%%%%%%%%%%%%%%%%%%%%%
%%%%%%%%%%%%%%%%%%%%%%%%%%%%%%%%%%%%%%%%
%%%%%%%%%%%%%%%%%%%%%%%%%%%%%%%%%%%%%%%%
%%%%%%%%%%%%%%%%%%%%%%%%%%%%%%%%%%%%%%%%

\section{D. Return current decrease in shearing runs} \label{app:A-RC-Decrease}

Let us consider a magnetic field line at a time $t$. We define $\rho$ as the distance 
between its two photospheric footpoints. $x_0$ and $y_0$ are the distance between 
both photospheric footpoints in the $x$ and $y$ direction respectively, such that
	\BA 		\label{eq:Eq-x-distance}
     		x_0 & = & \rho \sin \phi   \,  \\
		y_0 & = & \rho \cos \phi    \,,
		             \label{eq:Eq-y-distance}
     	\EA
where $\phi$ is the angle between the field line footpoints and the normal to the PIL 
at the photosphere. The variation of these distances during an infinitesimal time, 
$\mathrm{d} t$, is
	\BA 		\label{eq:Eq-dx-distance}
     		\mathrm{d} x_0 & = & \mathrm{d} \rho \sin \phi + \rho \cos \phi \ \mathrm{d} \phi  \,  \\
		\mathrm{d} y_0 & = & \mathrm{d} \rho \cos \phi - \rho \sin \phi \ \mathrm{d} \phi   \,.
		             \label{eq:Eq-dy-distance}
     	\EA
The photospheric motions being solely applied in the $x$ direction, it follows that 
$\mathrm{d} y_0 = 0$. Combining with \eqs{Eq-dx-distance}{Eq-dy-distance}, 
one obtains the following equations
	\BA 		\label{eq:Eq-drho-distance}
     		\mathrm{d} \rho & = & \sin \phi \ \mathrm{d} x_0   \,  \\
		\rho \mathrm{d} \phi & = & \cos \phi \ \mathrm{d} x_0   \,.
		             \label{eq:Eq-rhodphi-distance}
     	\EA
Then, we define $|\epsilon| = |\phi - \phi_0| \ll 1$. Replacing in 
\eqs{Eq-drho-distance}{Eq-rhodphi-distance} and expanding terms to a first order 
in $\epsilon$, one obtains
	\BA 		\label{eq:Eq-drho-DL}
     		\mathrm{d} \rho & \approx & \left( \sin \phi_0 + \epsilon \cos \phi_0 \right) \ \mathrm{d} x_0   \,  \\
		\rho \mathrm{d} \phi & \approx & \left(\cos \phi_0 - \epsilon \sin \phi_0 \right) \ \mathrm{d} x_0   \,.
		             \label{eq:Eq-rhodphi-DL}
     	\EA
When the footpoints segment is close to the normal to the PIL, $\phi_0 = 0$. It follows 
that $\mathrm{d} \rho \approx \epsilon \mathrm{d} x_0  \ll  \rho \mathrm{d} \phi \approx \mathrm{d} x_0$. 
The boundary driving essentially induces a rotation of the field line with regard to the vertical 
direction. In other words, it shears the magnetic field line, thus increasing its current density. 
On the contrary, when the footpoints segment is almost aligned with the PIL, $\phi_0 = \pi / 2$. 
Then, $\mathrm{d} \rho \approx \mathrm{d} x_0  \gg  \rho \mathrm{d} \phi \approx - \epsilon \mathrm{d} x_0$. 
The boundary driving essentially increases the distance between the field line footpoints. 
Since this distance is related to the field line length (larger footpoints distance implies larger 
field line), it follows that the photospheric motions essentially increase the length of the field line, 
hence reducing its current density \citep[\cf \eq{Eq-J-Cylindrical}; see also][]{Aulanier05}.

We then conclude that the continuous shearing of each magnetic field line can generate 
two evolutionary phases for their current density: (1) a first phase of increase with magnetic 
shear, and (2) a phase of decrease due to a fast elongation of the field lines. The shortest 
magnetic field lines (\ie with the smallest $y_0$) should be the first to be affected by 
the decrease of current due to their elongation because they align more rapidly with the PIL. 
For a given $x_0$, smaller values of $y_0$ induce larger values of $\tan \phi (t) = x_0 / y_0$, 
and hence, values of $\phi (t)$ closer to $\pi / 2$ (\ie the value for which the field line is aligned 
with the PIL).

\end{document}